\documentclass[11pt]{myArticle}       

\usepackage{amsmath,amssymb,amsfonts} 
\usepackage{graphicx}
\usepackage{epsfig}
\usepackage{color}                    
\usepackage{hyperref}                 
\usepackage{myFancyhdr}               
\usepackage{makeidx}                  
\usepackage{helvet}                   
\usepackage{setspace}                 
\usepackage{fancybox}                 
\usepackage{float}                    
\usepackage{xspace}                   
\usepackage{wasysym}                  
\usepackage[numbers,sort&compress]{natbib}
\usepackage{natbib}
\usepackage{hypernat}
\usepackage{wrapfig}

\usepackage{lineno}

\definecolor{mygray}{rgb}{0.3,0.32,0.35}
\definecolor{darkblue1}{rgb}{0,0,.2}
\definecolor{darkblue}{rgb}{0,0,.3}
\definecolor{darkred}{rgb}{0.5,0,0}
\pagecolor{white} 
\color{black}     
\hypersetup{breaklinks=true, 
            colorlinks=true, 
            linkcolor=darkblue1, 
            menucolor=darkblue1, 
            urlcolor=darkblue1,
            citecolor=darkblue1,
            pdftitle={The SM EW fit Revisited},
            pdfauthor={Gfitter Group},
            pdfsubject={The SM EW fit Revisited},
            pdfkeywords={},
            pdfproducer={Gfitter Group}
}
\bibstyle{plain}
\renewcommand{\headrulewidth}{0.5pt}    
\renewcommand{\footrulewidth}{0.5pt}    

\newcommand\allFontSize{\small}

\newenvironment{myquote}
               {\list{}{\leftmargin0cm}%
                \item\relax}
               {\endlist}

\usepackage[bf]{caption}

\newcommand\detailsSize{\allFontSize}
\newenvironment{details}%
{\begin{myquote}\vspace{-0.2cm}\detailsSize}{\end{myquote}\vspace{-0.2cm}}

\newfloat{codeexample}{H}{loc}
\floatname{codeexample}{\sf\allFontSize Code Example}

\newlength{\gfitterboxwidth}
\definecolor{DarkGray}{rgb}{0.4,0.42,0.45}
\definecolor{LightGray}{rgb}{0.97,0.98,0.98}

{\VerbatimEnvironment\normalsize
\setlength{\gfitterboxwidth}{\textwidth}\addtolength{\gfitterboxwidth}{-2.3\fboxsep}%
\begin{Sbox}\begin{minipage}[t]{\gfitterboxwidth}\begin{Verbatim}}%
{\end{Verbatim}\end{minipage}\end{Sbox}\vspace{1ex} \fcolorbox{DarkGray}{LightGray}{\TheSbox}}

\newfloat{option}{H}{loc}
\floatname{option}{\sf\allFontSize Option Table}

\usepackage{tabularx}
{\setlength{\gfitterboxwidth}{\textwidth}\addtolength{\gfitterboxwidth}{-2.3\fboxsep}%
\begin{Sbox}\begin{tabular*}{\gfitterboxwidth}{>{\rule[-0.5ex]{.0ex}{3.0ex}\tt\small}p{3cm}>{\tt\small}p{4.5cm}>{\small}p{5.7cm}}
&&\\[-0.7cm]
\rm\small Option  & \rm\small Values    & \rm\small Description\\[0.1cm]\hline
&&\\[-0.45cm]}%
{\\[-0.1cm]\end{tabular*}\end{Sbox}\vspace{1ex} \fbox{\TheSbox}}

{\setlength{\gfitterboxwidth}{\textwidth}\addtolength{\gfitterboxwidth}{-2.3\fboxsep}%
\begin{Sbox}\begin{tabular*}{\gfitterboxwidth}{>{\rule[-0.5ex]{.0ex}{3.0ex}\tt\small}p{3cm}>{\tt\small}p{3.5cm}>{\small}p{6.7cm}}
&&\\[-0.7cm]
\rm\small Option  & \rm\small Values    & \rm\small Description\\[0.1cm]\hline
&&\\[-0.45cm]}%
{\\[-0.1cm]\end{tabular*}\end{Sbox}\vspace{1ex} \fbox{\TheSbox}}

{\setlength{\gfitterboxwidth}{\textwidth}\addtolength{\gfitterboxwidth}{-2.3\fboxsep}%
\begin{Sbox}\begin{tabular*}{\gfitterboxwidth}{>{\rule[-0.5ex]{.0ex}{3.5ex}\tt\small}p{3.0cm}>{\tt\small}p{3.0cm}>{\small}p{7.2cm}}
&&\\[-0.7cm]
\rm\small Option  & \rm\small Values    & \rm\small Description\\[0.1cm]\hline
&&\\[-0.45cm]}%
{\\[-0.1cm]\end{tabular*}\end{Sbox}\vspace{1ex} \fbox{\TheSbox}}

{\setlength{\gfitterboxwidth}{\textwidth}\addtolength{\gfitterboxwidth}{-2.3\fboxsep}%
\begin{Sbox}\begin{tabular*}{\gfitterboxwidth}{>{\rule[-0.5ex]{.0ex}{3.0ex}\tt\small}p{2.5cm}>{\tt\small}p{3.5cm}>{\small}p{7.2cm}}
&&\\[-0.7cm]
\rm\small Option  & \rm\small Values    & \rm\small Description\\[0.1cm]\hline
&&\\[-0.45cm]}%
{\\[-0.1cm]\end{tabular*}\end{Sbox}\vspace{1ex} \fbox{\TheSbox}}

{\setlength{\gfitterboxwidth}{\textwidth}\addtolength{\gfitterboxwidth}{-2\fboxsep}%
\begin{Sbox}\begin{tabular*}{\gfitterboxwidth}{>{\rule[-0.5ex]{.0ex}{3.0ex}\tt\small}p{4.5cm}>{\tt\small}p{3.2cm}>{\small}p{5.5cm}}
&&\\[-0.7cm]
\rm\small Option  & \rm\small Values    & \rm\small Description\\[0.1cm]\hline
&&\\[-0.45cm]}%
{\\[-0.1cm]\end{tabular*}\end{Sbox}\vspace{1ex} \fbox{\TheSbox}}

{\setlength{\gfitterboxwidth}{\textwidth}\addtolength{\gfitterboxwidth}{-2\fboxsep}%
\begin{Sbox}\begin{tabular*}{\gfitterboxwidth}{>{\rule[-0.5ex]{.0ex}{3.0ex}\tt\small}p{4.0cm}>{\tt\small}p{3.0cm}>{\small}p{6.2cm}}
&&\\[-0.7cm]
\rm\small Option  & \rm\small Values    & \rm\small Description\\[0.1cm]\hline
&&\\[-0.45cm]}%
{\\[-0.1cm]\end{tabular*}\end{Sbox}\vspace{1ex} \fbox{\TheSbox}}

\newfloat{programs}{H}{loc}
\floatname{programs}{\sf Table}

\usepackage{tabularx}
{\setlength{\gfitterboxwidth}{\textwidth}\addtolength{\gfitterboxwidth}{-2\fboxsep}%
\begin{Sbox}\begin{tabular*}{\gfitterboxwidth}{>{\rule[-0.5ex]{.0ex}{3.0ex}\tt\small}p{4.1cm}>{\small}p{9.5cm}}
&\\[-0.7cm]
\rm\small Macro & \rm\small Description\\[0.1cm]\hline
&\\[-0.45cm]}%
{\\[-0.1cm]\end{tabular*}\end{Sbox}\vspace{1ex} \fbox{\TheSbox}}

\newcommand{\MSbar}{\ensuremath{\overline{\rm MS}}\xspace}
\mathchardef\Upsilon="7107
\def\Y#1S{\ensuremath{\Upsilon{(#1S)}}\xspace}

\newcommand{\mt}{\ensuremath{m_{t}}\xspace}
\newcommand{\MW}{\ensuremath{M_{W}}\xspace}
\newcommand{\MH}{\ensuremath{M_{H}}\xspace}
\newcommand{\as}{\ensuremath{\alpha_{\scriptscriptstyle S}}\xspace}

\newcommand{\asZ}{\ensuremath{\as(M_Z^2)}\xspace}

\renewcommand\l{\ell}

\newcommand{\GFe}[1]{{\ensuremath{G_{\scriptscriptstyle F}^{#1}}}\xspace}

\newcommand{\Kbar    }{\kern 0.2em\overline{\kern -0.2em K}{}\xspace}

\newcommand{\Kz      }{\ensuremath{K^0}\xspace}
\newcommand{\Kzb     }{\ensuremath{\Kbar^0}\xspace}
\newcommand{\KzKzb   }{\ensuremath{\Kz \kern -0.16em \Kzb}\xspace}
\newcommand{\Kp      }{\ensuremath{K^+}\xspace}
\newcommand{\Km      }{\ensuremath{K^-}\xspace}

\newcommand{\KpKm    }{\ensuremath{\Kp \kern -0.16em \Km}\xspace}

\newcommand\Dbar    {\kern 0.18em\overline{\kern -0.18em D}{}\xspace}

\newcommand\Bbar    {\kern 0.18em\overline{\kern -0.18em B}{}\xspace}

\newcommand\Bz      {\ensuremath{B^0}\xspace}

\newcommand\Bzb     {\ensuremath{\Bbar^0}\xspace}
\newcommand\Bu      {\ensuremath{B^+}\xspace}
\newcommand\Bub     {\ensuremath{B^-}\xspace}

\newcommand\BpBm    {\ensuremath{\Bu {\kern -0.16em \Bub}}\xspace}
\newcommand\Bs      {\ensuremath{B^0_{s}}\xspace}
\newcommand\Bsb     {\ensuremath{\Bbar^0_{s}}\xspace}

\newcommand\BzBzb   {\ensuremath{\Bz {\kern -0.16em \Bzb}}\xspace}
\newcommand\BszBszb {\ensuremath{\Bs {\kern -0.16em \Bsb}}\xspace}

\newcommand\deltatheo{\ensuremath{\delta_{\rm theo}}\xspace}

\newcommand\Rfit{{\em R}fit\xspace}

\newcommand{\ft}{\footnotesize}
\newcommand{\multic}{\multicolumn}

\newcommand{\Order}{\ensuremath{{\cal O}}\xspace}

\newcommand{\ee}{\ensuremath{e^+e^-}\xspace}

\newcommand{\tev}{\ensuremath{\mathrm{Te\kern -0.1em V}}\xspace}
\newcommand{\gev}{\ensuremath{\mathrm{Ge\kern -0.1em V}}\xspace}
\newcommand{\mev}{\ensuremath{\mathrm{Me\kern -0.1em V}}\xspace}
\newcommand{\kev}{\ensuremath{\mathrm{ke\kern -0.1em V}}\xspace}
\newcommand{\ev}{\ensuremath{\mathrm{e\kern -0.1em V}}\xspace}
\newcommand{\gevc}{\ensuremath{{\mathrm{Ge\kern -0.1em V\!/}c}}\xspace}
\newcommand{\mevc}{\ensuremath{{\mathrm{Me\kern -0.1em V\!/}c}}\xspace}
\newcommand{\gevcc}{\ensuremath{{\mathrm{Ge\kern -0.1em V\!/}c^2}}\xspace}
\newcommand{\mevcc}{\ensuremath{{\mathrm{Me\kern -0.1em V\!/}c^2}}\xspace}

\newcommand{\bei}{\begin{itemize}}
\newcommand{\eei}{\end{itemize}}
\newcommand{\beq}{\begin{equation}}
\newcommand{\eeq}{\end{equation}}
\newcommand{\beqn}{\begin{eqnarray}}
\newcommand{\eeqn}{\end{eqnarray}}
\newcommand{\beqns}{\begin{eqnarray*}}
\newcommand{\eeqns}{\end{eqnarray*}}
\newcommand{\bitm}{\begin{itemize}}
\newcommand{\eitm}{\end{itemize}}

\newcommand{\alphaMZ}{\ensuremath{\alpha(M_Z^2)}\xspace}

\newcommand{\dahadZf}{\ensuremath{\Delta\alpha_{\rm had}^{(5)}(M_Z^2)}\xspace}
\newcommand{\dalphaHadMZ}{\ensuremath{\Delta\alpha_{\rm had}^{(5)}(M_Z^2)}\xspace}

\newcommand\rs{\raisebox{1.5ex}[-1.5ex]}

\newcommand{\ST}{$S,\,T$\xspace}
\newcommand{\STU}{$S,\,T,\,U$\xspace}
\def\@citex[#1]#2{\if@filesw\immediate\write\@auxout{\string\citation{#2}}\fi
  \@tempcnta\z@\@tempcntb\m@ne\def\@citea{}\@cite{\@for\@citeb:=#2\do
    {\@ifundefined
       {b@\@citeb}{\@citeo\@tempcntb\m@ne\@citea
        \def\@citea{,\penalty\@m\ }{\bf ?}\@warning
       {Citation `\@citeb' on page \thepage \space undefined}}%
    {\setbox\z@\hbox{\global\@tempcntc0\csname b@\@citeb\endcsname\relax}%
     \ifnum\@tempcntc=\z@ \@citeo\@tempcntb\m@ne
       \@citea\def\@citea{,\penalty\@m}
       \hbox{\csname b@\@citeb\endcsname}%
     \else
      \advance\@tempcntb\@ne
      \ifnum\@tempcntb=\@tempcntc
      \else\advance\@tempcntb\m@ne\@citeo
      \@tempcnta\@tempcntc\@tempcntb\@tempcntc\fi\fi}}\@citeo}{#1}}

\def\@citeo{\ifnum\@tempcnta>\@tempcntb\else\@citea
  \def\@citea{,\penalty\@m}%
  \ifnum\@tempcnta=\@tempcntb\the\@tempcnta\else
   {\advance\@tempcnta\@ne\ifnum\@tempcnta=\@tempcntb \else
\def\@citea{--}\fi
    \advance\@tempcnta\m@ne\the\@tempcnta\@citea\the\@tempcntb}\fi\fi}

\xspace

\newcommand\mini{{\rm min}}

\newcommand\ChiMin{\ensuremath{\chi^2_{\mini}}\xspace}
\newcommand\DeltaChi{\ensuremath{\Delta\chi^2}\xspace}

\newcommand{\seffsf}[1]{\sin\!^2\theta^{#1}_{{\rm eff}}}
\newcommand{\ceffsf}[1]{\cos\!^2\theta^{#1}_{{\rm eff}}}
\newcommand{\sinfeff}{\ensuremath{\seffsf{f}}\xspace}
\newcommand{\sinleff}{\ensuremath{\seffsf{\ell}}\xspace}
\newcommand{\cosleff}{\ensuremath{\ceffsf{\ell}}\xspace}

\newcommand{\mc}{\ensuremath{\overline{m}_c}\xspace}
\newcommand{\mb}{\ensuremath{\overline{m}_b}\xspace}

\newcommand{\ChiMinVal}    {\ensuremath{17.8}\xspace}

\newcommand{\NDFVal}     	{\ensuremath{14}\xspace}

\newcommand{\SParam}     	{\ensuremath{0.05\pm 0.11}\xspace}
\newcommand{\TParam}     	{\ensuremath{0.09\pm 0.13}\xspace}
\newcommand{\UParam}     	{\ensuremath{0.01\pm 0.11}\xspace}

\newcommand{\STParamCor}	{\ensuremath{+0.90}\xspace}
\newcommand{\SUParamCor}	{\ensuremath{-0.59}\xspace}
\newcommand{\TUParamCor}	{\ensuremath{-0.83}\xspace}

\newcommand{\SParamNU}     {\ensuremath{0.06\pm 0.09}\xspace}
\newcommand{\TParamNU}     {\ensuremath{0.10\pm 0.07}\xspace}
\newcommand{\STParamCorNo} {\ensuremath{+0.91}\xspace}

\makeindex
\begin{document}


%
%
\pagenumbering{arabic}
{\small
\color{mygray}
\begin{flushright}
{\sf\em DESY-14-124} \\
{\sf\em \today} \\
\def\UrlFont{\sf\em}
\url{http://cern.ch/gfitter} 
\end{flushright}
}
\def\UrlFont{\rm}

\vspace{1.3cm}


{\sf\LARGE\bfseries
The global electroweak fit at NNLO and \\[0.15cm] prospects for the LHC and ILC
}

\vspace{2.0cm}

{\large \em 
  The Gfitter Group \\[0.2cm]
}
{\large
  M.~Baak$^{a}$, J.~C\'uth$^{b}$, J.~Haller$^{c}$, A.~Hoecker$^{a}$, R.~Kogler$^{c}$, 
  K.~M\"onig$^{d}$, M.~Schott$^{b}$, J.~Stelzer$^{e}$
}

\vspace{0.2cm}

\begin{details}
  $^{a}$CERN, Geneva, Switzerland \\
  $^{b}$Institut f\"ur Physik, Universit\"at Mainz, Germany\\
  $^{c}$Institut f\"ur Experimentalphysik, Universit\"at Hamburg, Germany\\
  $^{d}$DESY, Hamburg and Zeuthen, Germany \\ 
  $^{e}$Department of Physics and Astronomy, Michigan State University, East Lansing, USA
\end{details}

\vspace{2.0cm}

\begin{details} {\sf\bfseries Abstract} 
  --- For a long time, global fits of the electroweak sector of the
  Standard Model (SM) have been used to exploit measurements of
  electroweak precision observables at lepton colliders (LEP, SLC),
  together with measurements at hadron colliders (Tevatron, LHC), and
  accurate theoretical predictions at multi-loop level, to constrain
  free parameters  of the SM, such as the Higgs and top masses.  Today, all
  fundamental SM parameters entering these fits
  are experimentally determined, including 
  information on the Higgs couplings, and the global fits are used as
  powerful tools to assess the validity of the theory and to constrain
  scenarios for new physics. Future measurements at the Large Hadron
  Collider (LHC) and the International Linear Collider (ILC) 
  promise to improve the experimental precision of 
  key observables used in the fits.
  This paper presents updated electroweak fit results using newest NNLO
  theoretical predictions, and prospects for the LHC and ILC. The 
  impact of experimental and theoretical uncertainties is analysed in 
  detail. We compare constraints from the electroweak fit on the Higgs couplings
  with direct LHC measurements, and examine present and future prospects
  of these constraints using a model with modified couplings of the Higgs 
  boson to fermions and bosons.

\end{details}

\thispagestyle{empty}

\newpage
%
%
\section{Introduction}
\label{sec:intro}

Global fits of the Standard Model (SM) have traditionally combined
electroweak precision observables with accurate theoretical
predictions to constrain the top quark and Higgs boson
masses~\cite{LEPEWWG, Erler:2010wa,Flacher:2008zq,Baak:2011ze,ref:1306.4644}. 
The discovery of a scalar boson at the Large Hadron Collider (LHC)~\cite{ATLASHiggs,CMSHiggs},
with mass $M_{H}$ around $125\;$GeV, provides an impressive confirmation of the light Higgs  
prediction derived from these fits. 
Assuming the new boson to be the SM Higgs boson and inserting the measured
mass into the fit overconstrains the electroweak sector of the SM. 
Key electroweak observables such as the $W$ boson mass, $M_W$, and the
effective weak mixing angle for charged and neutral leptons and light quarks, 
$\sinfeff$, can thus be predicted with a precision exceeding that of
the direct measurements~\cite{Baak:2012kk}.
These observables become sensitive probes of new physics~\cite{Eberhardt:2012np}
limited in part by the accuracy of the theoretical calculations.

Recently, full fermionic two-loop calculations have become available
for the partial widths and branching ratios of the $Z$ boson~\cite{Freitas:2014hra}. 
These new calculations improve the theoretical precision and also allow for a 
more meaningful estimate 
of the theoretical uncertainties due to missing higher perturbative orders.
In this paper we present an update of the global electroweak fit performed at the two-loop 
level,\footnote{The decay width of the $W$ boson, $\Gamma_W$, is only known to one-loop precision.
However, this measurement being of insufficient precision has a negligible impact 
on the result of the electroweak fit.} 
including a detailed assessment of the impact of the remaining theoretical uncertainties.

Measurements performed at future colliders will increase the experimental precision of these 
and other electroweak  observables, such as the top quark mass, $m_t$.  
The coming years should also lead to progress in the calculation of multi-loop 
corrections to these observables, as well as to an improved determination of the hadronic
contribution to the  fine-structure constant evaluated at the $Z$ boson mass scale, 
$\Delta \alpha_{\rm had}(M_Z^2)$. 

In this paper, the latest results of the global electroweak fit are compared with 
the expectations for the Phase-1 LHC\footnote{This 
corresponds to a scenario with 
$\int Ldt=300\;{\rm fb}^{-1}$ at $\sqrt{s}=14$\;TeV, before the high luminosity upgrade.}
and the International Linear Collider (ILC) with GigaZ
option,\footnote{GigaZ: the operation of the ILC at
  lower energies like the $Z$ pole or the $WW$ threshold allows for
  high-statistics precision measurements of several electroweak observables.
  At the $Z$ pole the physics at LEP and SLC can
  be revisited with the data collected during a few days. Several
  billion $Z$ bosons can be produced within a few 
  months~\cite{ILCTDR13}. In comparison: in the seven years 
  that LEP operated at the $Z$ peak it produced around 
  17 million $Z$ bosons in its four interaction points;
  SLD studied about 600 thousand $Z$ bosons produced with 
  a polarised beam~\cite{ALEPH:2005ema}.} 
henceforth denoted ILC/GigaZ~\cite{ILCTDR13}.  
For each scenario we analyse the impact of the assigned experimental and theoretical uncertainties. 
By exploiting contributions from radiative corrections, the global
electroweak fit is also used to determine the couplings of the
Higgs boson to gauge bosons using the formalism of the \STU
parameters.  
We combine the constraints on the Higgs 
couplings in a popular benchmark model with LHC measurements 
of the signal strength in various channels.
We also study the prospects for these constraints.

The paper is organised as follows. 
An update of the global electroweak fit including the recent theoretical improvements
is presented in Section~\ref{sec:smfit}. 
Section~\ref{sec:prospectsewfit} discusses
the extrapolated uncertainties of key input observables for the LHC and
ILC/GigaZ scenarios, the fit prospects, and a detailed
analysis of the impact of all sources of systematic uncertainties.
The status and prospects for the determinations of Higgs couplings from the electroweak fit 
are reported in Section~\ref{sec:higgscouplings}.

\section{Update of the global electroweak fit}
\label{sec:smfit}

In the following we present an update of the global electroweak fit at the $Z$-mass scale. 
The relevant observables, the data treatment and statistical framework are described 
in Refs.~\cite{Flacher:2008zq, Baak:2011ze}. 
We use the recent calculations of the $Z$ boson partial widths and branching ratios at
the electroweak two-loop order~\cite{Freitas:2014hra}. 
These provide for the first time a consistent set of calculations at next-to-next-to leading 
order (NNLO) for all relevant input observables, together with the  
two-loop calculations of the $W$ mass and the effective weak mixing angle.

\subsection*{SM predictions and theoretical uncertainties}

The following theoretical predictions of the SM observables are used.
\begin{itemize}
\item The effective weak mixing angle, \sinfeff, has been calculated using corrections
up to the full two-loop order $\mathcal{O}(\alpha \alpha_s)$ and 
$\mathcal{O}(\alpha^2)$~\cite{Awramik:2004ge, Awramik:2006uz}. 
In addition, partial three-loop and four-loop terms have been included 
at order $\mathcal{O}(\alpha \alpha_s^2)$~\cite{Avdeev:1994db, Chetyrkin:1995ix, Chetyrkin:1995js}, 
$\mathcal{O}(\alpha_t^2 \alpha_s)$, $\mathcal{O}(\alpha_t^3)$~\cite{vanderBij:2000cg, Faisst:2003px} and 
$\mathcal{O}(\alpha_t \alpha_s^3)$~\cite{Schroder:2005db, Chetyrkin:2006bj, Boughezal:2006xk}, where $\alpha_t = \alpha m_t^2$. 
These calculations have been included in the parametrisation provided 
in~\cite{Awramik:2004ge, Awramik:2006uz}. 
For bottom quarks the calculation from~\cite{Awramik:2008gi} is used,
which includes corrections of the same order together with
additional vertex corrections from top-quark propagators.
\item The mass of the $W$ boson, $M_W$, has been calculated 
to the same orders of electroweak and QCD corrections as \sinfeff. 
We use the parametrisation of the full two-loop result~\cite{Awramik:2003rn}.
New in this paper is the inclusion of four-loop QCD corrections 
$\mathcal{O}(\alpha_t \alpha_s^3)$~\cite{Schroder:2005db, Chetyrkin:2006bj, Boughezal:2006xk}, 
which result in a shift of the predicted $M_W$ by about $-2.2$\;\mev 
in the on-shell renormalisation scheme for $m_t$.
The exact value of the shift depends on the parameter settings used.
\item Full fermionic two-loop corrections $\mathcal{O}(\alpha^2)$ for 
the partial widths and branching ratios of the $Z$ boson have recently become 
available~\cite{Freitas:2014hra}.\footnote{These 
calculations do not include diagrams with closed boson loops
at two-loop order. These are expected to give small corrections 
compared to diagrams with closed fermion loops and
are therefore only considered in the estimate of the theoretical uncertainty.
}
The parametrisation formulas provided include also  higher order terms to 
match the perturbative order of the calculations of \sinfeff and $M_W$. 
These calculations amend previous two-loop 
predictions of the total $Z$ width, $\Gamma_Z$,  the hadronic peak
cross section, $\sigma^0_{\rm had}$~\cite{Freitas:2013dpa}, and the partial 
decay width of the $Z$ boson into $b\bar{b}$, $R^0_b$~\cite{Freitas:2012sy}.
\item The dominant contributions from final-state QED and QCD radiation
are included in the calculations through factorisable radiator functions $\mathcal{R_{V,A}}$,
which are known up to $\mathcal{O}(\alpha_s^4)$ for massless final-state
quarks, $\mathcal{O}(\alpha_s^3)$ for massive quarks~\cite{Chetyrkin:1996ia, Baikov:2008jh, Baikov:2012er}, 
and $\mathcal{O}(\alpha^2)$ for contributions with closed fermion loops~\cite{Kataev:1992dg}.
Non-factorisable vertex contributions of order $\mathcal{O}(\alpha \alpha_s)$~\cite{Czarnecki:1996ei, Harlander:1997zb} 
are also accounted for.
\item The width of the $W$ boson, $\Gamma_W$, is known up to one electroweak 
loop order. We use the parametrisation given in~\cite{Cho:2011rk},
which is sufficient given the limited experimental precision. 

\end{itemize}

In summary, the changes with respect to our previous publication~\cite{Baak:2012kk} are 
the addition of the $\mathcal{O}(\alpha_t \alpha_s^3)$ QCD correction to $M_W$ 
and the two-loop calculations of the partial $Z$ widths.
The latter calculations also yield an updated result for $R^0_b$, including non-factorisable 
$\mathcal{O}(\alpha\alpha_s)$ and $\mathcal{O}(\alpha_s^4)$ corrections.
For the calculations of the SM predictions we use the computer code employed
previously~\cite{Baak:2012kk}, with the corresponding updates.

\begin{table}[tb]
\begin{center}
\setlength{\tabcolsep}{0.5cm}
\begin{tabular}{ll | ll} 
\hline\noalign{\smallskip}
$\deltatheo M_W$ & 4\;\mev                                       &  $\deltatheo\Gamma_{u,c}$ & 0.12\;\mev \\
$\deltatheo \sinfeff$   & $4.7\cdot 10^{-5}$                  &  $\deltatheo\Gamma_{b}$   & 0.21\;\mev \\
$\deltatheo\Gamma_{e, \mu, \tau}$ & 0.012\;\mev     &  $\deltatheo \sigma^0_{\rm had}$     & 6\;pb  \\
$\deltatheo\Gamma_{\nu}$ & 0.014\;\mev                  &  $\deltatheo \mathcal{R}_{V,A}$        &  $\sim \mathcal{O}(\alpha_s^4)$    \\
$\deltatheo\Gamma_{d,s}$ & 0.09\;\mev                    & $\deltatheo m_t$        & 0.5\;\gev \\
\noalign{\smallskip}\hline\noalign{\smallskip}
\end{tabular}
\end{center}
\vspace{-0.4cm}
\caption{
Theory uncertainties taken into account in the global electroweak fit. See text for details. 
\label{tab:theo_unc}
}
\end{table}

The theoretical uncertainties from unknown higher-order contributions have 
been estimated by assuming that the perturbation series follow a geometric 
growth~\cite{Awramik:2003rn, Awramik:2006uz, Freitas:2014hra}. 
The resulting uncertainties for $M_W$, \sinfeff, $\sigma^0_{\rm had}$ and all decay widths, 
$\Gamma_f$, for the decay $Z \to f\bar{f}$, are listed in Table~\ref{tab:theo_unc}.

For $M_W$ and $\sinfeff$ they arise from three dominant sources of unknown higher-order corrections: $\Order(\alpha^2\as)$
terms beyond the known contribution of $\Order(\GFe{2} \as \mt^4)$,
$\Order(\alpha^3)$ electroweak three-loop corrections, and
$\Order(\as^3)$ QCD terms. 
Summing quadratically the relevant
uncertainty estimates amounts to the overall theoretical uncertainties
$\deltatheo M_W=4\;\mev$~\cite{Awramik:2003rn} and
$\deltatheo\sinfeff=4.7 \cdot 10^{-5}$~\cite{Awramik:2006uz}.

The leading theoretical uncertainties on the predicted $Z$ decay widths and $\sigma^0_{\rm had}$
come from missing two-loop electroweak bosonic $\Order(\alpha^2)$ contributions, three-loop terms 
of order $\Order(\alpha^3)$, $\Order(\alpha^2\as)$ and $\Order(\alpha\as^2)$, and 
$\Order(\alpha\as^3)$ corrections beyond the leading $m_t^n$ terms. 
The resulting uncertainties $\deltatheo\Gamma_f$ 
are between $0.012$ and $0.21\;\mev$ (see Table~\ref{tab:theo_unc}). 
The theoretical uncertainty $\deltatheo \sigma^0_{\rm had}$ amounts to~6\;pb. 

Uncertainties due to unknown higher order contributions to 
the radiator functions $\mathcal{R}_{V,A}$ 
have been estimated by varying the $\mathcal{O}(\alpha_s^4)$ terms for the 
massless and massive quark contributions by factors of $0$ to $2$. 
The uncertainty due to the singlet vector contribution was found to be negligible. 

We assign an additional theoretical uncertainty to the value of $m_t$ from 
hadron collider measurements due to the ambiguity in the kinematic 
top-mass definition~\cite{Hoang:2008yj,Hoang:2008xm,Buckley:2011ms,Moch:2014tta},
the colour structure of the fragmentation process~\cite{Skands:2007zg,Wicke:2008iz},
and the perturbative relation between pole and \MSbar mass 
currently known to three-loop order~\cite{Chetyrkin:1999qi, Melnikov:2000qh}.
The first uncertainty is difficult to assess. Estimates range from 0.25 to 
0.9\;\gev or higher~\cite{top13mangano,Moch:2014tta}. 
Systematic effects on \mt due to mis-modeling of the colour
reconnection in the fragmentation process, initial and final state
radiation, and the kinematics of the $b$-quark are partly considered as 
uncertainties by the experiments. They were also studied in a dedicated 
measurement by CMS~\cite{CMS:2012ixa} where no significant trends 
between the measurements under different conditions were observed. 
Finally, estimates of the missing higher order perturbative correction to the relation 
between the pole and \MSbar top mass range from 0.2 to 0.3\;\gev uncertainty~\cite{Hoang:2000yr}.  
The nominal value of the combined \mt uncertainty is set here to 0.5\;\gev. 
The impact of this uncertainty is studied below for values between 0 and 1.5\;\gev.

Our previous publications~\cite{Flacher:2008zq, Baak:2011ze, Baak:2012kk}
employed the \Rfit scheme, characterised by uniform likelihoods for the two 
theoretical nuisance parameters $\deltatheo M_W$ and $\deltatheo \sinfeff$ used. 
It corresponds to a linear addition of theoretical and experimental uncertainties.
In this analysis, with the ten theoretical nuisance parameters listed in Table~\ref{tab:theo_unc}, 
we use Gaussian constraints to stabilise the fit convergence. 
The Gaussian treatment modifies the relative impact of theoretical uncertainties. 
The fit constraints become tighter at low significance and looser at high significance, 
which -- to keep in mind -- is important when interpreting the results.

\subsection*{Experimental input}

A detailed list of all the observables, their values and uncertainties used in the fit, 
is given in the first two columns of Table~\ref{tab:present_results}.
The input data to the fit consist of measurements at the $Z$ pole by the LEP 
and SLD collaborations~\cite{ALEPH:2005ema}, the world average values
for the running quark masses~\cite{Beringer:1900zz}, 
of $M_W$ and $\Gamma_W$~\cite{Beringer:1900zz}, 
and an up-to-date determination of the five-quark hadronic  
vacuum polarisation contribution to \alphaMZ, \dahadZf~\cite{Davier:2010nc}.
For the mass of the top quark we use the latest average from the
direct measurements by the LHC and Tevatron experiments~\cite{ATLAS:2014wva}
with the additional 0.5\;\gev theoretical uncertainty as discussed above. 
The mass of the Higgs boson $M_H$ is taken to be the average of the  
new results by ATLAS~\cite{Aad:2014aba} ($125.4 \pm 0.4$\;\gev) 
and CMS~\cite{CMS-PAS-HIG-14-009} ($125.0 \pm 0.3$\;\gev), 
$125.14 \pm 0.24$\;\gev.

\subsection*{Results of the SM fit}

\begin{table}
\setlength{\tabcolsep}{0.0pc}
{\small
\begin{tabular*}{\textwidth}{@{\extracolsep{\fill}}lccccc} 
\hline\noalign{\smallskip}
& & Free &  & \multic{1}{c}{w/o exp. input} & \multic{1}{c}{w/o exp. input}   \\[-0.1cm]
\rs{Parameter} & \rs{Input value} & in fit & \rs{Fit Result} & \multic{1}{c}{in line} & \multic{1}{c}{in line, no theo. unc} \\
\noalign{\smallskip}\hline\noalign{\smallskip}
$M_{H}$ {\ft [GeV]}$^{(\circ)}$ &  $125.14\pm0.24$ & yes & $125.14\pm0.24$ & $93^{+25}_{-21}$ & $93^{+24}_{-20}$\\
\noalign{\smallskip}\hline\noalign{\smallskip}
$M_{W}$ {\ft [GeV]} &  $80.385\pm0.015$ & -- &  $80.364\pm0.007$ &  $80.358\pm0.008$ &  $80.358\pm0.006$\\
$\Gamma_{W}$ {\ft [GeV]} &  $2.085\pm0.042$ & -- &  $2.091\pm0.001$ &  $2.091\pm0.001$ &  $2.091\pm0.001$\\
\noalign{\smallskip}\hline\noalign{\smallskip}
$M_{Z}$ {\ft [GeV]} &  $91.1875\pm0.0021$ & yes &  $91.1880\pm0.0021$ &  $91.200\pm0.011$ &  $91.2000\pm0.010$\\
$\Gamma_{Z}$ {\ft [GeV]} &  $2.4952\pm0.0023$ & -- &  $2.4950\pm0.0014$ &  $2.4946\pm0.0016$ &  $2.4945\pm0.0016$\\
$\sigma_{\rm had}^{0}$ {\ft [nb]} &  $41.540\pm0.037$ & -- &  $41.484\pm0.015$ &  $41.475\pm0.016$ &  $41.474\pm0.015$\\
$R^{0}_{\l}$ &  $20.767\pm0.025$ & -- &  $20.743\pm0.017$ &  $20.722\pm0.026$ &  $20.721\pm0.026$\\
$A_{\rm FB}^{0,\l}$ &  $0.0171\pm0.0010$ & -- &  $0.01626\pm0.0001$ &  $0.01625\pm0.0001$ &  $0.01625\pm0.0001$\\
 $A_\ell$ $^{(\star)}$  & $0.1499\pm0.0018$ & --  & $0.1472\pm0.0005$ & $0.1472\pm0.0005$ & $0.1472\pm0.0004$\\
$\sinleff(Q_{\rm FB})$ &  $0.2324\pm0.0012$ & -- &  $0.23150\pm0.00006$ &  $0.23149\pm0.00007$ &  $0.23150\pm0.00005$\\
$A_{c}$ &  $0.670\pm0.027$ & -- &  $0.6680\pm0.00022$ &  $0.6680\pm0.00022$ &  $0.6680\pm0.00016$\\
$A_{b}$ &  $0.923\pm0.020$ & -- &  $0.93463\pm0.00004$ &  $0.93463\pm0.00004$ &  $0.93463\pm0.00003$\\
$A_{\rm FB}^{0,c}$ &  $0.0707\pm0.0035$ & -- &  $0.0738\pm0.0003$ &  $0.0738\pm0.0003$ &  $0.0738\pm0.0002$\\
$A_{\rm FB}^{0,b}$ &  $0.0992\pm0.0016$ & -- &  $0.1032\pm0.0004$ &  $0.1034\pm0.0004$ &  $0.1033\pm0.0003$\\
$R^{0}_{c}$ &  $0.1721\pm0.0030$ & -- &  $0.17226^{\,+0.00009}_{\,-0.00008}$ &  $0.17226\pm0.00008$ &  $0.17226\pm0.00006$\\
$R^{0}_{b}$ &  $0.21629\pm0.00066$ & -- &  $0.21578\pm0.00011$ &  $0.21577\pm0.00011$ &  $0.21577\pm0.00004$\\
\noalign{\smallskip}\hline\noalign{\smallskip}
$\mc$ {\ft [GeV]} &  $1.27^{\,+0.07}_{\,-0.11}$ & yes &  $1.27^{\,+0.07}_{\,-0.11}$ & --  & -- \\
$\mb$ {\ft [GeV]} &  $4.20^{\,+0.17}_{\,-0.07}$ & yes &  $4.20^{\,+0.17}_{\,-0.07}$ & --  & -- \\
$m_{t}$ {\ft [GeV]} &  $173.34\pm0.76$ & yes &  $173.81\pm0.85$$^{(\bigtriangledown)}$ &  $177.0^{\,+2.3}_{\,-2.4}$$^{(\bigtriangledown)}$ &  $177.0\pm2.3$\\
$\dalphaHadMZ$$^{(\dag\bigtriangleup)}$ &  $2757\pm  10$ & yes & $2756\pm  10$ & $2723\pm  44$ & $2722\pm  42$\\
$\alpha_{s}(M_{Z}^{2})$ & -- & yes &  $0.1196\pm0.0030$ &  $0.1196\pm0.0030$ &  $0.1196\pm0.0028$\\
\noalign{\smallskip}\hline
\noalign{\smallskip}
\end{tabular*}
{\ft
\begin{flushleft}
\vspace{-0.3cm}
$^{(\circ)}$Average of the ATLAS~\cite{Aad:2014aba} and CMS~\cite{CMS-PAS-HIG-14-009} measurements assuming no 
correlation of the systematic uncertainties.\\
$^{(\star)}$Average of the LEP and SLD $A_\ell$ measurements~\cite{ALEPH:2005ema}, 
used as two measurements in the fit.\\
$^{(\bigtriangledown)}$The theoretical top mass uncertainty of 0.5 GeV is excluded.\\ 
$^{(\dag)}$In units of $10^{-5}$.\\
$^{(\bigtriangleup)}$Rescaled due to $\alpha_s$ dependence.
\end{flushleft}
} }
\caption{Input values and fit results for the observables used in the global electroweak fit. 
The first and second columns list respectively the observables/parameters used in the fit, 
and their experimental values or phenomenological estimates (see text for references). 
The third column indicates whether a parameter is floating in the fit. The fourth column 
quotes the results of the fit including all experimental data. In the fifth column the 
fit results are given without using the corresponding experimental or phenomenological 
estimate in the given row (indirect determination). The last column shows for illustration 
the result using the 
same fit setup as in the fifth column, but ignoring all theoretical uncertainties. 
The nuisance parameters that are used to parameterise theoretical uncertainties 
are given in Table~\ref{tab:theo_unc}.
\label{tab:present_results}
}
\end{table}

The fit of the electroweak theory to all input data from Table~\ref{tab:present_results}, 
including the theoretical uncertainties from Table~\ref{tab:theo_unc},
converges at a global minimum value of $\ChiMin= \ChiMinVal$, obtained for \NDFVal\ degrees of freedom. 
Using pseudo experiments and the statistical method described in~\cite{Flacher:2008zq} 
we find a $p$-value for the SM to describe the data of $0.21$ (corresponding to $0.8\sigma$ one-sided significance).
The improved goodness-of-fit compared to earlier results~\cite{Baak:2012kk} 
comes mostly from the corrected calculation of $R^0_b$~\cite{Freitas:2012sy}, 
which decreases the previously reported discrepancy of $R^0_b$ between the global fit and the measurement  
from a pull value of $-2.4\sigma$ down to $-0.8\sigma$
(consistent with the one-loop calculation of $R^0_b$). 
The impact of this change on the other fit parameters is small.
The new two-loop calculations of the $Z$ partial widths decrease the value of \ChiMin by 0.2, 
whereas the $\mathcal{O}(\alpha_t \alpha_s^3)$ four-loop QCD corrections  to $M_W$ increase
\ChiMin by 0.4 units.

\begin{figure}[tbp]
\begin{center}
\includegraphics[width=7.8cm]{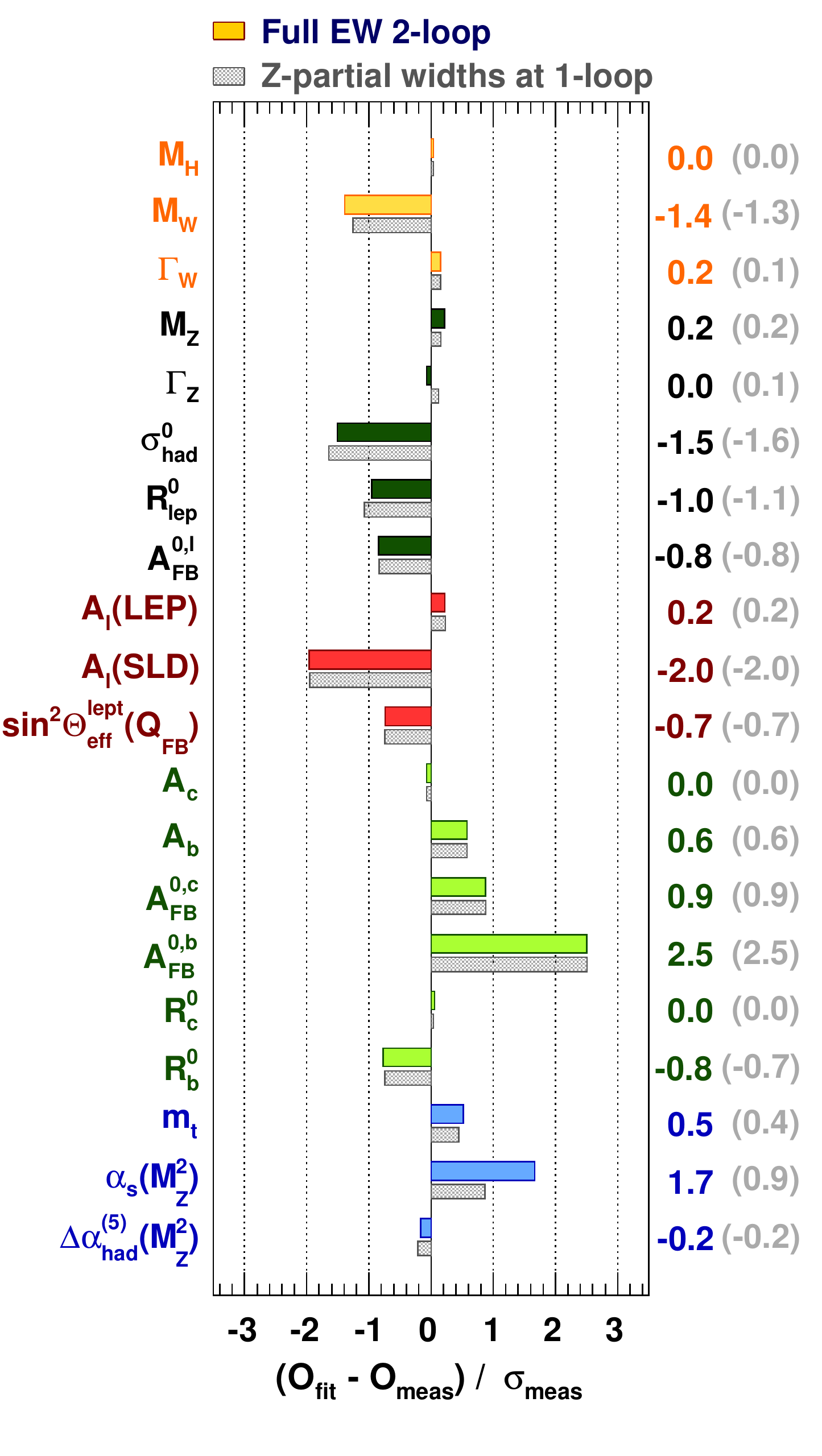}  \hspace{0.3cm}
\includegraphics[width=7.8cm]{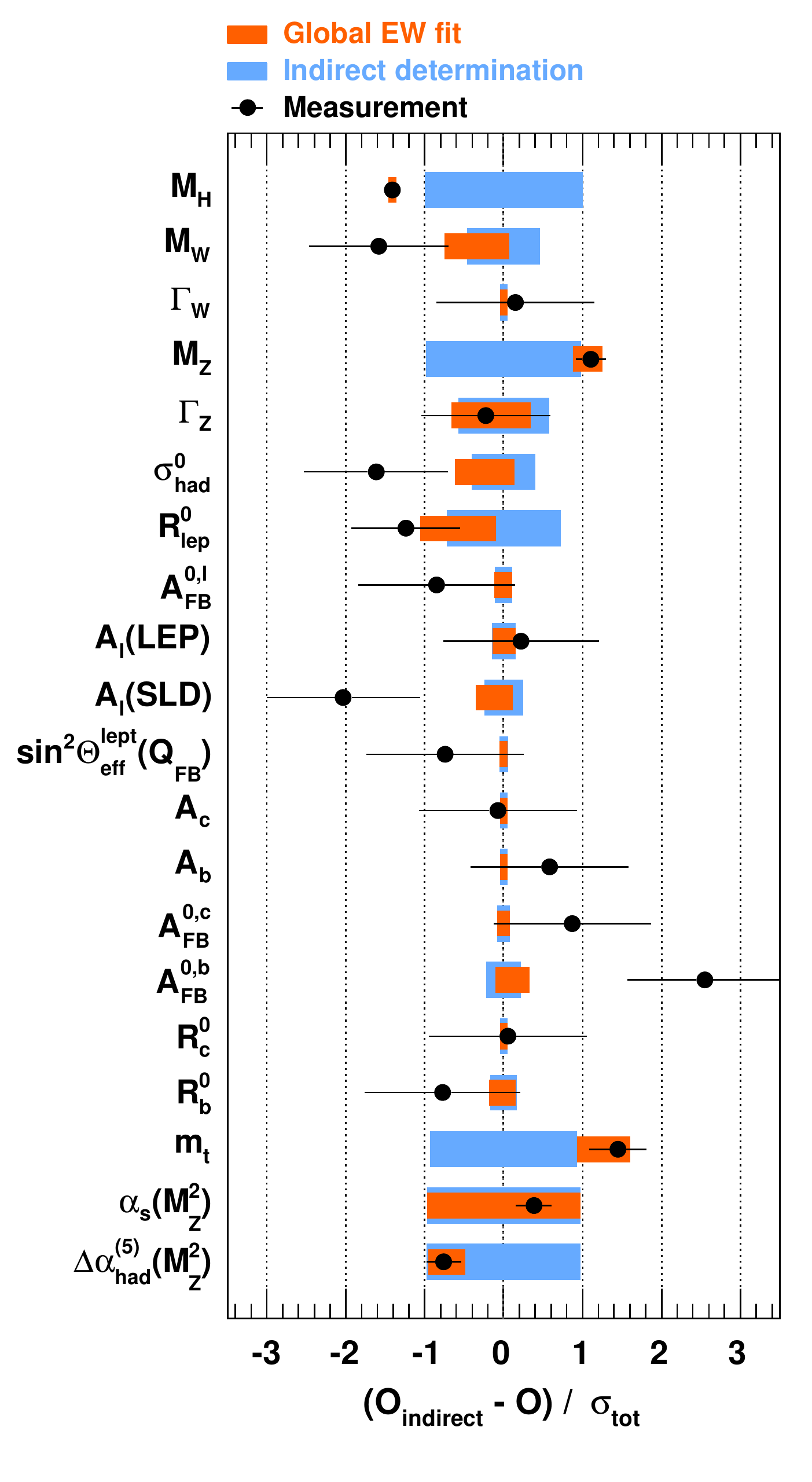} 
\end{center}
\vspace{-0.5cm}
\caption{
{\textit Left:} Comparison of the fit results with the direct measurements in units of the experimental uncertainty. 
The fit results are compared between the scenario using the two-loop calculations of the $Z$ partial widths 
with the four-loop $\mathcal{O}(\alpha_t \alpha_s^3)$ correction to $M_W$ (colour, top bars), 
and the one-loop calculation used in a previous publication~\cite{Baak:2011ze} (shaded gray, bottom bars).
{\textit Right:} Comparison of the fit results with the indirect determination in units of the total uncertainty, 
defined as the uncertainty of the direct measurement and that of the indirect determination added in quadrature. 
The indirect determination of an observable corresponds to a fit without using the corresponding direct constraint from the measurement.
\label{fig:pullplot}
}
\end{figure}

The results of the full fit for each fit parameter and observable are
given in the fourth column of Table~\ref{tab:present_results},
together with the uncertainties estimated from their $\Delta\chi^2=1$ profiles.  
The fifth column in Table~\ref{tab:present_results} gives
the results obtained without using in the fit 
the experimental measurement corresponding to that row.
A more detailed discussion of these indirect determinations 
for several key observables is given in
Section~\ref{sec:fitresults}. 
The last column in Table~\ref{tab:present_results} corresponds to the fits of the previous column 
but ignoring, for the purpose of illustration, all theoretical uncertainties.
In this case the global fit converges at a slightly increased 
minimum value of $\ChiMin / {\rm ndf}= 18.2/\NDFVal$.

The result of the fit is summarised in Fig.~\ref{fig:pullplot}. 
The plot on the left shows a comparison of the global fit results (fourth column of Tab.~\ref{tab:present_results}) 
with the direct measurements (first column of Tab.~\ref{tab:present_results})
in units of the measurement uncertainty.
Also shown is the impact of the two-loop result for the
$Z$ partial widths and the $\mathcal{O}(\alpha_t \alpha_s^3)$ correction to $M_W$, 
compared to the calculations previously used\footnote{With 
the exception of $R^0_b$, which was previously
  taken from~\cite{Freitas:2012sy} and was later corrected. For
  this comparison the one-loop result~\cite{Cho:2011rk} is used.
}~\cite{Baak:2012kk}.  
The right-hand panel of Fig.~\ref{fig:pullplot} displays the comparison of 
both the global fit result and the direct measurements 
with the indirect determination (fifth column of
Tab.~\ref{tab:present_results}) for each observable
in units of the total uncertainty, defined as the uncertainty 
of the direct measurement and the indirect determination added in quadrature.
Note that in the case of $\alpha_{s}(M_{Z}^{2})$ the direct
measurement displayed is the world average value~\cite{Beringer:1900zz},
which is otherwise not used in the fit.

The availability of the two-loop corrections to the $Z$ partial widths and $\sigma^0_{\rm had}$ 
allows the determination of $\alpha_{s}(M_{Z}^{2})$ to full NNLO and partial NNNLO level. 
We find
\beqn
\alpha_{s}(M_{Z}^{2}) &=& 0.1196 \pm 0.0028_{\mathrm{\,exp}}    \pm 0.0006_{\delta_{\rm theo} \mathcal{R}_{V,A}}    \pm 0.0006_{\delta_{\rm theo} \Gamma_{i}}    \pm 0.0002_{\delta_{\rm theo} \sigma^0_{\rm had}}   \nonumber \\ 
                     &=& 0.1196 \pm 0.0030_{\mathrm{\,tot}}\;,
\eeqn
where the theoretical uncertainties due to missing higher order contributions 
are significantly larger than previously estimated~\cite{Baak:2012kk}. 
This is largely due to the variation of the full $\mathcal{O}(\alpha_s^4)$ terms 
in the radiator functions, and to the uncertainties on the $Z$ partial widths and
$\sigma^0_{\rm had}$, not assigned before.

The fit indirectly determines the $W$ mass to be
\beqn
  M_W &=& 80.3584 
          \pm 0.0046_{m_t} \pm 0.0030_{\delta_{\rm theo} m_t} \pm 0.0026_{M_Z} \pm 0.0018_{\Delta\alpha_{\rm had}} \nonumber \\ 
      & & \phantom{80.3584} 
          \pm  0.0020_{\as} \pm 0.0001_{M_H} \pm 0.0040_{\delta_{\rm theo} M_W}\;{\rm GeV}\,, \nonumber \\[0.2cm]
      &=& 80.358 \pm 0.008_{\rm tot} \;{\rm GeV} \,.
\label{eq:mw}
\eeqn
providing a result which exceeds the precision of the direct measurement.
The different uncertainty contributions originate from the uncertainties on the input 
values of the fit, as quoted in the second column in Table~\ref{tab:present_results}. 
Simple error-propagation is applied to evaluate their impact on the prediction of
$M_{W}$.
At present, the largest uncertainties are 
due to $m_t$, both experimental and theoretical, 
followed by the theory and $M_Z$ uncertainties.

Likewise, the indirect determination of the effective leptonic weak mixing angle, $\sinleff$, gives
\beqn
  \sinleff &=& 0.231488 
                 \pm 0.000024_{m_t} \pm 0.000016_{\delta_{\rm theo} m_t} \pm 0.000015_{M_Z} \pm 0.000035_{\Delta\alpha_{\rm had}} \nonumber \\
           & & \phantom{0.231496}
                 \pm 0.000010_{\as} \pm 0.000001_{M_H} \pm 0.000047_{\delta_{\rm theo} \sinfeff} \,, \nonumber \\[0.2cm]
      &=& 0.23149 \pm 0.00007_{\rm tot} \;,
\label{eq:sin2t}
\eeqn
where the largest uncertainty is theoretical followed by the
uncertainties on $\dalphaHadMZ$ and $m_t$.

\begin{figure}[tbp]
\begin{center}
\includegraphics[width=0.73\textwidth]{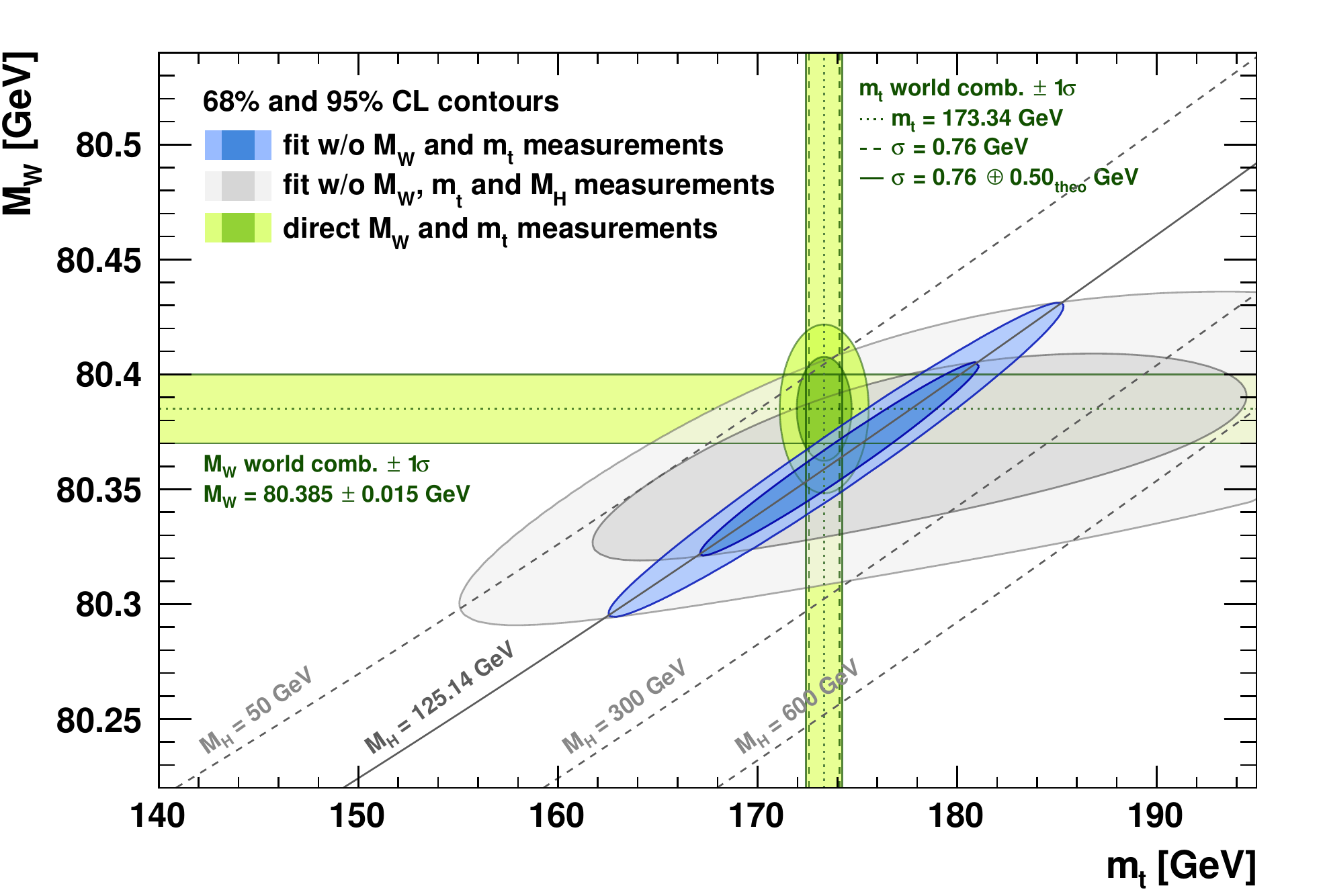}

\vspace{0.3cm}
\includegraphics[width=0.73\textwidth]{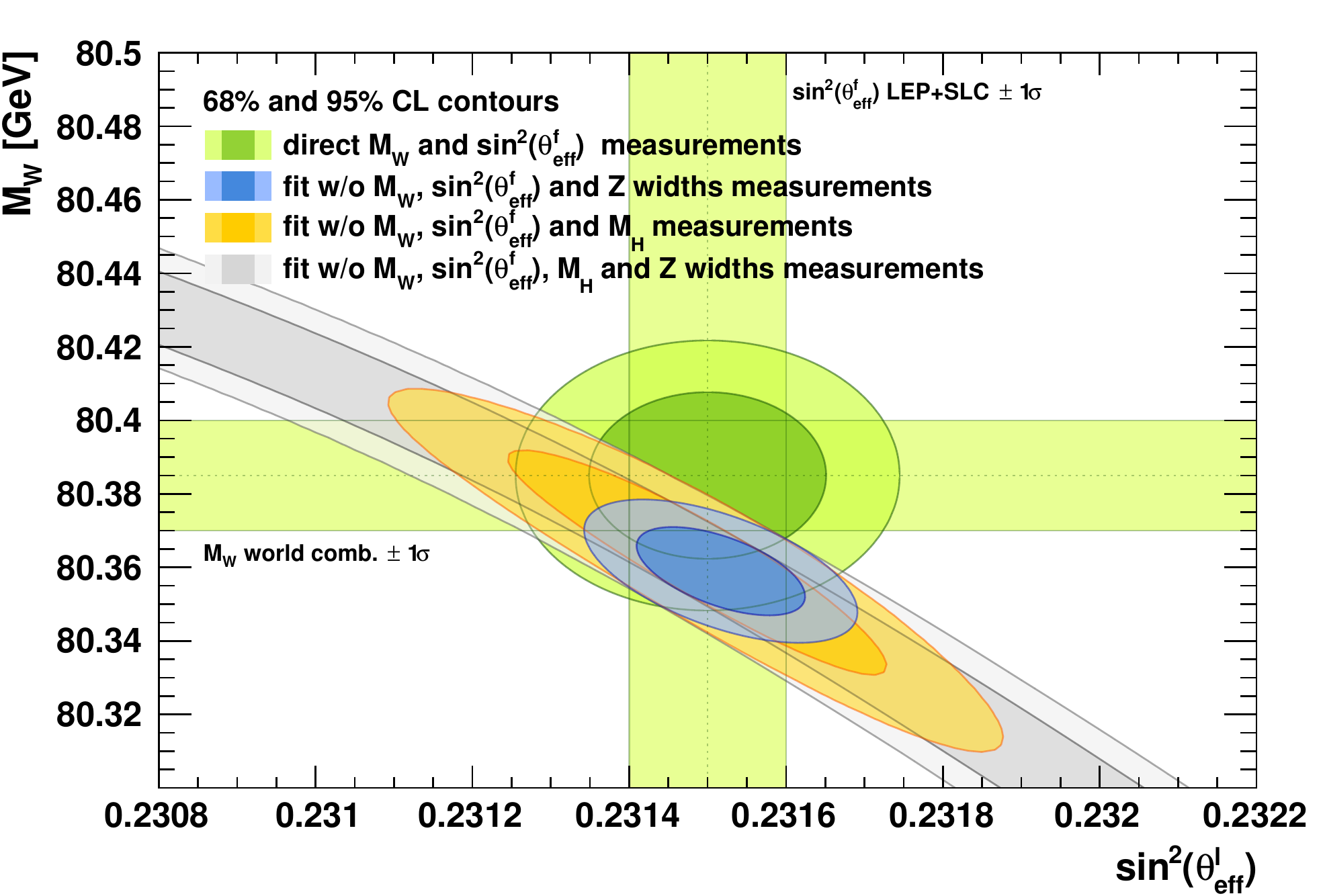}
\end{center}
\vspace{-0.1cm}
  \caption[]{ Contours at 68\% and 95\% CL obtained from scans
    of $M_W$ versus $m_t$ (top) and $M_W$ versus $\sinleff$ (bottom),
    for the fit including \MH (blue) and excluding \MH (grey), 
    as compared to the direct measurements (vertical and horizontal 
    green bands and ellipses). 
    The theoretical uncertainty of 0.5 GeV is added to the direct top mass measurement. 
    In both figures, the corresponding direct measurements are excluded from the fit.
    In the case of $\sinleff$, all partial and full $Z$ width measurements are excluded 
    as well (except in case of the orange prediction), besides the asymmetry measurements.}
\label{fig:wvss2t}
\end{figure}

An important consistency test of the SM is the simultaneous indirect determination of 
\mt and \MW.
A scan of the confidence level (CL) profile of \MW versus \mt  
is shown in Fig.~\ref{fig:wvss2t} (top) for the scenarios 
where the direct \MH measurement is included in the fit (blue) or not (grey).
Both contours agree with the direct measurements (green bands and ellipse for 
two degrees of freedom).
The bottom panel of Fig.~\ref{fig:wvss2t} displays the corresponding CL
profile for the observable pair $\sinleff$ and $M_W$. 
The coloured ellipses
indicate: green for the direct measurements; grey for the electroweak fit
without using \MW, $\sinfeff$, \MH and the $Z$ width measurements;
orange for the fit without using \MW, $\sinfeff$ and \MH; blue for the fit 
without \MW, $\sinfeff$ and the $Z$ width measurements.
For both figures the observed agreement demonstrates the consistency of the SM.

\begin{figure}[tbp]
\begin{center}
\includegraphics[width=0.73\textwidth]{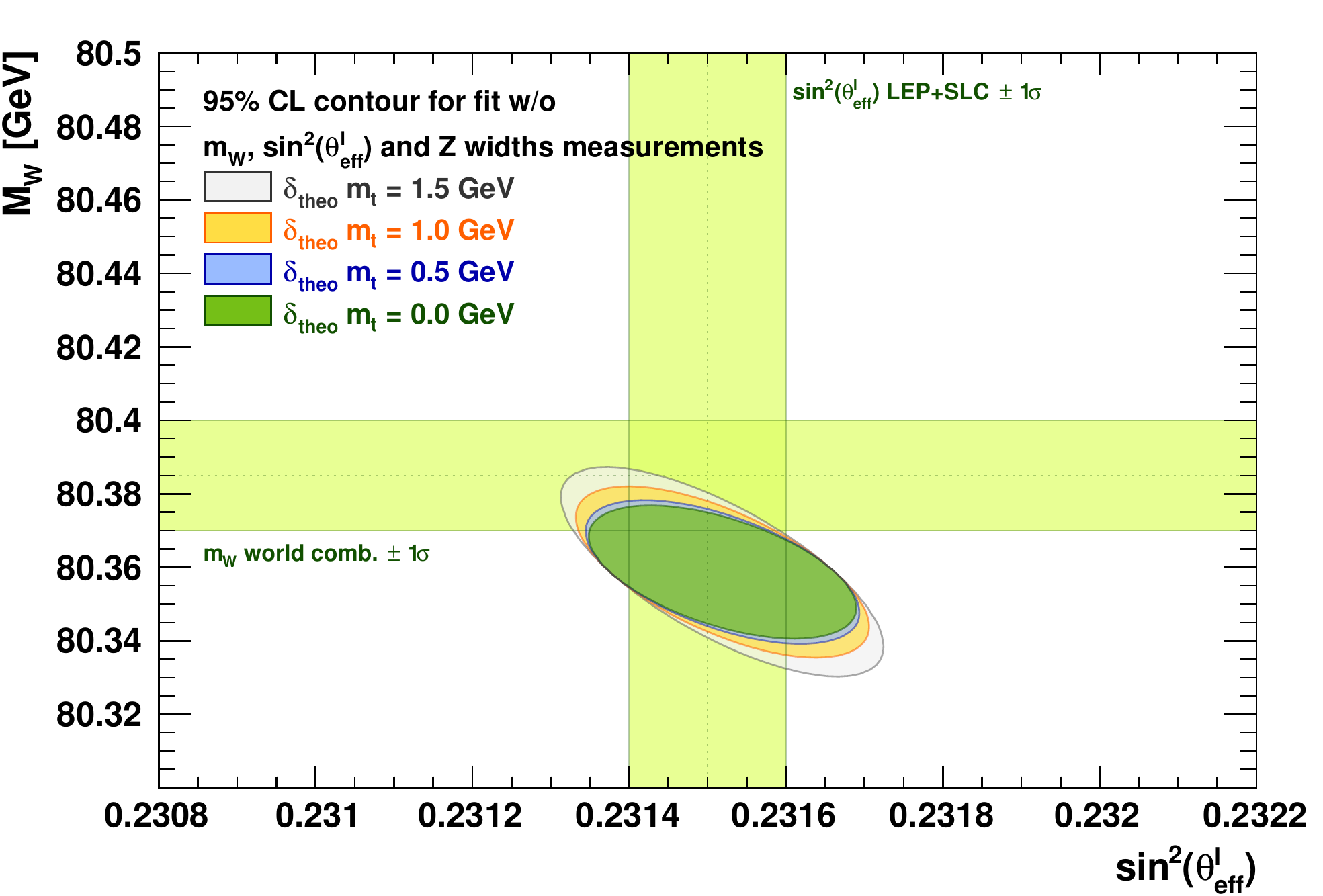}
\end{center}
\vspace{-0.1cm}
  \caption[]{ Contours at 95\% CL obtained from scans of $M_W$ versus $\sinleff$,
    with the top mass theoretical uncertainty varied between 0 and 1.5\;\gev in steps of 0.5\;\gev, 
    as compared to the direct measurements (vertical and horizontal green bands).
    The corresponding direct measurements are excluded from the fit.}
\label{fig:wvss2t2}
\end{figure}

Fig.~\ref{fig:wvss2t2} shows CL profiles for the observable pair $\sinleff$ and $M_W$,
but with the theoretical uncertainty on the top mass varied between 0 and 1.5\;\gev, in steps of 0.5\;\gev. 
It underlines that a better assessment of the theoretical \mt uncertainty 
is of relevance for the fit.

\subsection*{Oblique parameters}

If the new physics scale is significantly higher than the electroweak scale, 
new physics effects from virtual particles in loops are expected to contribute 
predominantly through vacuum polarization corrections to the electroweak precision 
observables. 
These terms are traditionally denoted {\it oblique corrections}
and are conveniently parametrised by the three self-energy parameters
\STU~\cite{Peskin:1990zt,Peskin:1991sw}, which are defined to vanish in the SM. 
The $S$ and $T$ parameters absorb
possible new physics contributions to the neutral and to the difference between 
neutral and charged weak currents, respectively. The $U$ parameter is only sensitive 
to changes in the mass and width of the $W$ boson. It is very
small in most new physics models and therefore often set to zero.

Constraints on the \STU parameters can be derived from the global
electroweak fit by calculating the difference of the oblique
corrections as determined from the experimental data and the
corrections obtained from an SM reference point (with fixed reference
values of $m_t$ and $M_H$). 
With this definition significantly
non-zero \STU parameters represent an unambiguous indication of new
physics. 

\begin{figure}[tp]
\begin{center}
\includegraphics[width=0.73\textwidth]{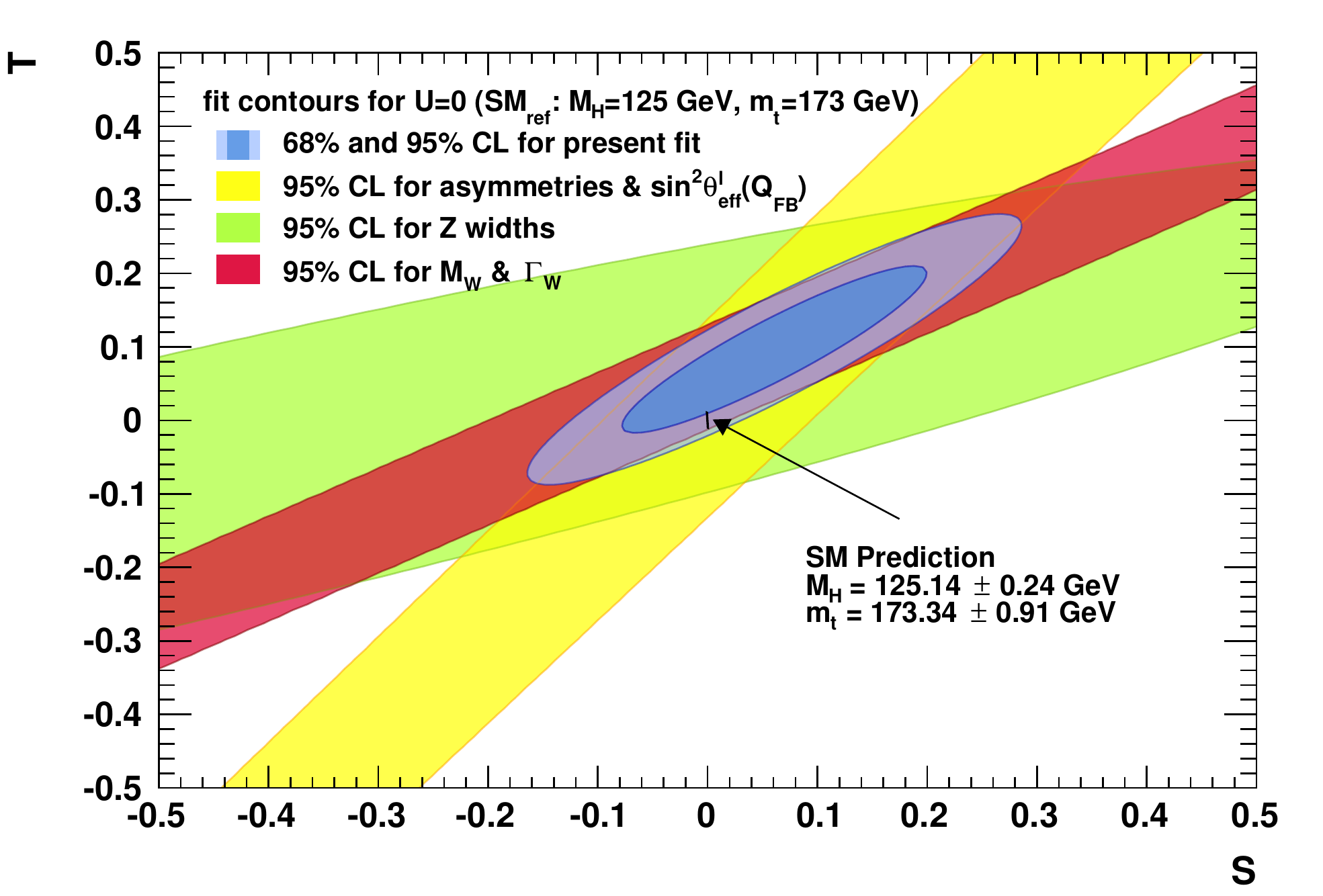}
\end{center}
\vspace{-0.1cm}
\caption[]{Constraints on the oblique parameters
  $S$ and $T$, with the $U$ parameter fixed to zero, using all observables (blue). 
  Individual constraints are shown from the asymmetry measurements (yellow),
  the Z partial and total widths (green) and $W$ mass and width (red),
  with confidence levels drawn for one degree of freedom.
  The SM prediction within uncertainties is indicated by the 
  thin black stroke.}
\label{fig:STU}
\end{figure}

For the studies presented here we use the SM reference as 
$M_{H,{\rm  ref}}=125$\;GeV and $m_{t,{\rm ref}}=173$\;GeV.
We find
\beq
  S= \SParam\:, \hspace{0.5cm}
  T= \TParam\:, \hspace{0.5cm}
  U=\UParam\:,
\eeq 
with correlation coefficients of $\STParamCor$ between $S$ and $T$, $\SUParamCor$ ($\TUParamCor $) between $S$
and $U$ ($T$ and $U$). 
Fixing $U=0$ one obtains $S|_{U=0}= \SParamNU$ and $T|_{U=0}= \TParamNU$, with a correlation coefficient of
$\STParamCorNo$. 
The constraints on $S$ and $T$ for a fixed 
value of $U=0$ are shown in Fig.~\ref{fig:STU}. 
The propagation of the current experimental uncertainties in $M_H$ and
$m_t$ upon the SM prediction is illustrated by the small black area at
about $S=T=0$.

\section{Prospects of the electroweak fit with the LHC and ILC/GigaZ}
\label{sec:prospectsewfit}

We use a simplified set of input observables to study the prospects of the 
electroweak fit for the Phase-1 LHC and the ILC/GigaZ. 
The measurements of the $Z$ pole asymmetry observables are summarised in 
a single value of the effective weak mixing angle. The measurement of $R^0_\l$ is the 
only partial decay width that enters the fit to constrain $\as$.
This simplified fit setup leads in some cases to reduced constraints on 
observables as can be seen by comparing the uncertainties of the present 
scenarios between the last column of Table~\ref{tab:present_results} and 
the fifth column of Table~\ref{tab:results_best}.
The central values of the  observables are adjusted to the values
predicted by the current best fit giving
a fully consistent set of SM observables.\footnote{
  The following central values are used for the future 
  scenarios: $M_H = 125.0$\;GeV, $\dalphaHadMZ = 2755.4 \cdot 10^{-5}$, 
  $M_Z = 91.1879$\;GeV, $m_t = 173.81$\;GeV, $M_W = 80.363$\;GeV, 
  $\sinleff = 0.231492$ and $R^0_\l = 20.743$. 
  See Table~\ref{tab:results_best} 
  for the corresponding uncertainties.}

\subsection*{Experimental and theoretical improvements}

For the LHC, with a large dataset and sufficient time to understand and 
improve systematic uncertainties, we assume the following scenario. 
\begin{itemize}
\item For $m_H$ an uncertainty of 100\;MeV is assumed, although the
  experiments are expected to exceed this precision using, for
  example, Higgs decays to four muons. Whatever uncertainty used is
  irrelevant for the fits discussed.
\item A precision of $10\;$MeV on $M_W$ may be achievable for
  the final combination of Tevatron measurements~\cite{Bozzi:2011ww}.
  Assuming improvements in the uncertainties due to parton distribution
  functions, the modelling of the lepton
  transverse momentum and a reduction of experimental uncertainties,
  we expect that a combined precision of $8\;$MeV may be in reach 
  for a combination of the LHC, Tevatron and LEP results.
\item Given the present combined $m_t$ uncertainty
  of $0.76\;$GeV~\cite{ATLAS:2014wva},  
  we assume an ultimate experimental precision of $0.6\;$GeV 
  as a long-term prospect. 
  As discussed earlier, an additional theoretical uncertainty of 0.5\:\gev is assigned.
  For the future LHC scenario, with further theoretical studies on the top mass ambiguity,
  additional high-statistics tests of top quark decay kinematics,
  and a possible perturbative four-loop relation between pole and \MSbar 
  mass~\cite{Hoang:2000yr,Melnikov:2000qh}, this  
  uncertainty is assumed to be reduced to 0.25\:\gev.
\end{itemize}

For the ILC/GigaZ we assume the following benchmark 
uncertainties\footnote{
An improvement in the $M_Z$ precision from
  currently 2.1\;MeV to 1.6\;MeV is suggested in~\cite{ILCTDR13}.  Such
  a measurement would require the knowledge of the absolute ILC beam
  energy with a precision of $10^{-5}$. Since the technical
  feasibility of such a precision is still uncertain, we do not
  yet include it in the fit.}.
\begin{itemize}
\item A precision of 5\;MeV is assumed for $M_W$, obtained from cross section
  measurements at and above the $WW$ production threshold~\cite{ILCTDR13}.
\item Scans of the $t\bar{t}$ production threshold are expected to
  yield an experimental precision on the top quark mass of
  approximately $30$\;MeV~\cite{ILCTDR13, Hoang:2000yr}. 
  The conversion of the threshold to an $\overline{\rm MS}$ mass using
  perturbative QCD adds an estimated uncertainty of
  100\;MeV~\cite{ILCTDR13, Hoang:2000yr, Chetyrkin:1999qi}.
\item Measurements of the weak left-right asymmetry $A_{\rm LR}$ from
  leptonic and hadronic $Z$ decays are expected to yield a precision of 
  $1.3\cdot 10^{-5}$ for  $\sinfeff$~\cite{ILCTDR13}, improving 
  the present measurement combination by more than a factor of ten.
\item The partial decay width of the $Z$ boson, $R_\l^0$, is assumed to 
  be measured with a precision of $4\cdot 10^{-3}$, improving the current 
measurement~\cite{Hawkings:1999ac} by a factor of more than six.
\end{itemize}

For both future scenarios we assume that the uncertainty in
$\dalphaHadMZ$ will reduce from currently $10\cdot 10^{-5}$ down to
$4.7\cdot 10^{-5}$.
The improvement is expected due to updated $\ee\to{\rm hadrons}$ cross section
measurements below the charm threshold from the completion of ongoing 
BABAR and VEPP-2000 analyses, improved charmonium resonance data from BES-III, 
and a better knowledge of \as from reliable Lattice QCD predictions~\cite{Davier:2012}.

The present and projected experimental 
uncertainties for the observables 
used in the simplified electroweak fit are summarised in the left columns of 
Table~\ref{tab:results_best}. 

To match the experimental precision significant theoretical progress is required.
Leaving aside the ambiguity in $m_t$, the presently most important theoretical 
uncertainties affecting the fit are those related to the 
predictions of $M_W$ and $\sinfeff$. 
For the future scenarios, we assume that the present uncertainties of 
$\deltatheo M_W=4\;\mev$ and $\deltatheo\sinfeff=4.7 \cdot 10^{-5}$
reduce to $1\;$MeV and $10^{-5}$, respectively. 
This reduction will require ambitious three-loop electroweak calculations. 
The leading theoretical uncertainties on the partial $Z$ decay widths, $\sigma^0_{\rm had}$, 
and the radiator functions play a smaller role in the present fit.
For the future scenarios the uncertainty estimates given in Table~\ref{tab:theo_unc} 
are assumed to be reduced by a factor of four, similar to the uncertainties on 
$M_W$ and $\sinfeff$. 

\begin{table}[t]
\setlength{\tabcolsep}{0.0pc}
{\normalsize                                                                                   
\begin{tabular*}{\textwidth}{@{\extracolsep{\fill}}lcccccc}                                                                                                                                 
\hline\noalign{\smallskip}
                                   & \multicolumn{3}{c}{Experimental input [{\footnotesize $\pm 1 \sigma_{\rm exp}$]}   } &  
                                   \multicolumn{3}{c}{Indirect determination [{\footnotesize $\pm 1\sigma_{\rm exp}, \, \pm 1\sigma_{\rm theo}$}]} \\
Parameter               &   Present &   LHC  &   ILC/GigaZ                   &  Present      & LHC   & ILC/GigaZ \\
\noalign{\smallskip}\hline\noalign{\smallskip}
$ M_{H}$        {\footnotesize[GeV]}              & $0.4$ & $<0.1$ & $<0.1$ & $_{-26}^{+31} \, , \; _{-8}^{+10}$   & $_{-18}^{+20}\, , \; _{-3.2}^{+3.9}$  & $_{-6.6}^{+6.9}\, , \; _{-2.3}^{+2.5}$ \\
$ M_{W}$        {\footnotesize [MeV]}              & $15 $ & $ 8$  & $5  $ &  $ 6.0 , \; 5.0$        & $ 5.2 ,\; 1.8$     & $ 1.9 ,\; 1.3$        \\
$ M_{Z}$        {\footnotesize [MeV]}              & $2.1$ & $2.1$ & $2.1$ &  $ 11 ,\; 4$      & $  7.0 ,\; 1.4$       & $ 2.6 ,\; 1.0$       \\
$ m_{t}$        {\footnotesize [GeV]}              & $0.8$ & $0.6$ & $0.1$ & $ 2.4 ,\; 0.6$        & $  1.5 ,\; 0.2$  & $ 0.7 ,\; 0.2$ \\
$\sinleff$   {\footnotesize $[10^{-5}]$} & $16$  & $16$  & $1.3$ & $ 4.5 ,\; 4.9 $       & $ 2.8 ,\; 1.1$      & $ 2.0 ,\; 1.0$        \\
$ \Delta \alpha^{5}_{\rm had}(M_Z^2)$ {\footnotesize $[10^{-5}]$} & $10$  & $4.7$ & $4.7$ & $ 42 ,\; 13$   & $ 36 ,\; 6$  & $  5.6 ,\; 3.0$   \\
$ R_l^0             $ {\footnotesize $[10^{-3}]$} & $25$        & $25$        & $4$         & --                & --               & --                \\
$ \asZ$            {\footnotesize $[10^{-4}]$} & --         & --         & --         & $  40,\; 10$         & $ 39 ,\; 7$      & $ 6.4 ,\; 6.9$           \\
\noalign{\smallskip}\hline\noalign{\smallskip}
$ S|_{U=0}$           &  --     &   --  &   --                    & $  0.094 ,\; 0.027 $ &  $  0.086 ,\; 0.006 $   & $  0.017 ,\; 0.006 $     \\
$ T|_{U=0}$           &  --     &   --  &   --                    & $  0.083 ,\; 0.023 $ &  $  0.064 ,\; 0.005 $   & $  0.022 ,\; 0.005 $     \\
\noalign{\smallskip}\hline\noalign{\smallskip}
$ \kappa_{V}$ {\footnotesize ($\lambda\;=\; 3$\,TeV)}            & $ 0.05$   &  $ 0.03$   &  $ 0.01$                 & $ 0.02$ & $ 0.02$ & $ 0.01$ \\        
\noalign{\smallskip}\hline
\noalign{\smallskip}
\end{tabular*} \\
}
\caption{Current and extrapolated future uncertainties in the input observables 
  (left), and the precision obtained for the fit prediction (right). Where two uncertainties are
  given, the first is experimental and the second theoretical. The value 
  of $\as(M_Z^2)$ is not used directly as input in the fit. 
  The uncertainty in the direct $M_H$ measurements is not relevant for the fit and 
  therefore not quoted. For all indirect determinations shown 
  (including the present $M_H$ determination) the assumed central values of 
  the input measurements have been adjusted to obtain a common fit value of 
  $M_H=125$\;GeV. 
  The simplified fit setup used to derive the numbers in this table leads in 
  some cases to reduced constraints on 
  observables as can be seen by comparing the uncertainties of the present 
  scenarios (fifth column) with the last column of Table~\ref{tab:present_results}.
  See text for more details.}
\label{tab:results_best} 
\end{table}

\subsection*{Expected fit performance}\label{sec:fitresults}

The numerical 1$\sigma$ uncertainties of the indirect observable determinations 
are given for the present fit as well as the LHC and ILC/GigaZ scenarios
in the right-hand columns of Table~\ref{tab:results_best}.
Experimental and theoretical uncertainties are quoted separately.

\begin{figure}[p!]
\begin{center}
\includegraphics[width=0.55\textwidth]{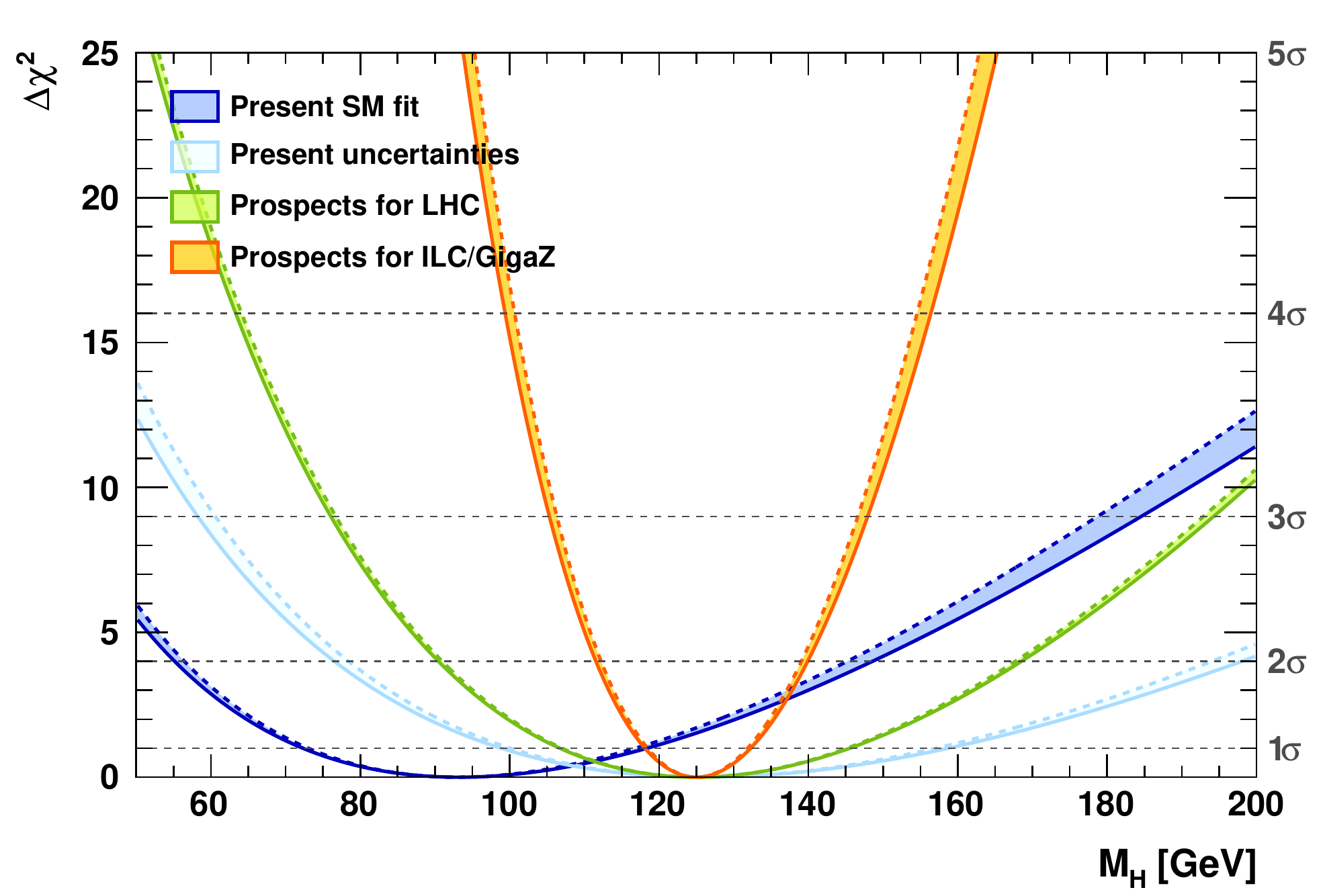}
\includegraphics[width=0.55\textwidth]{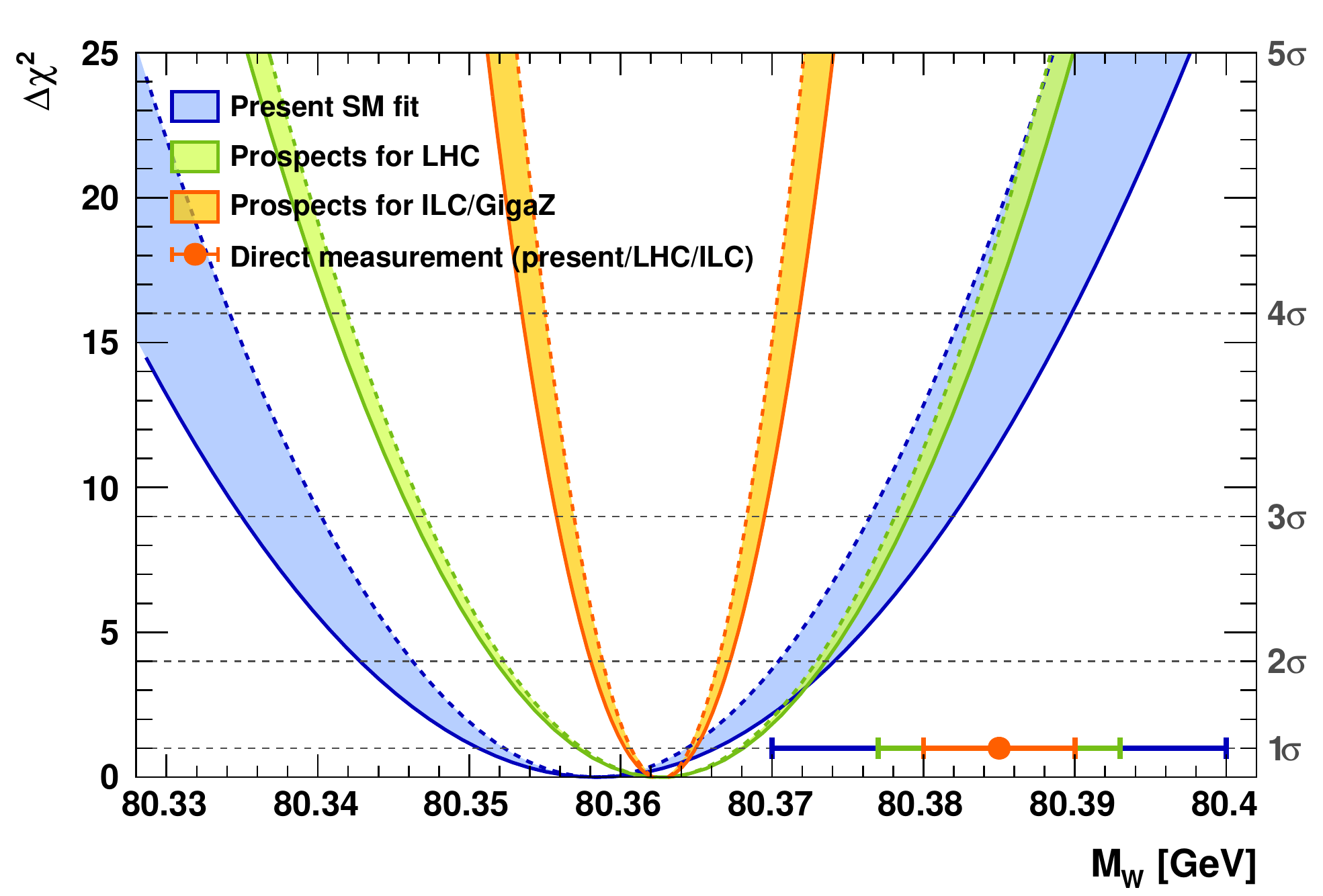}
\includegraphics[width=0.55\textwidth]{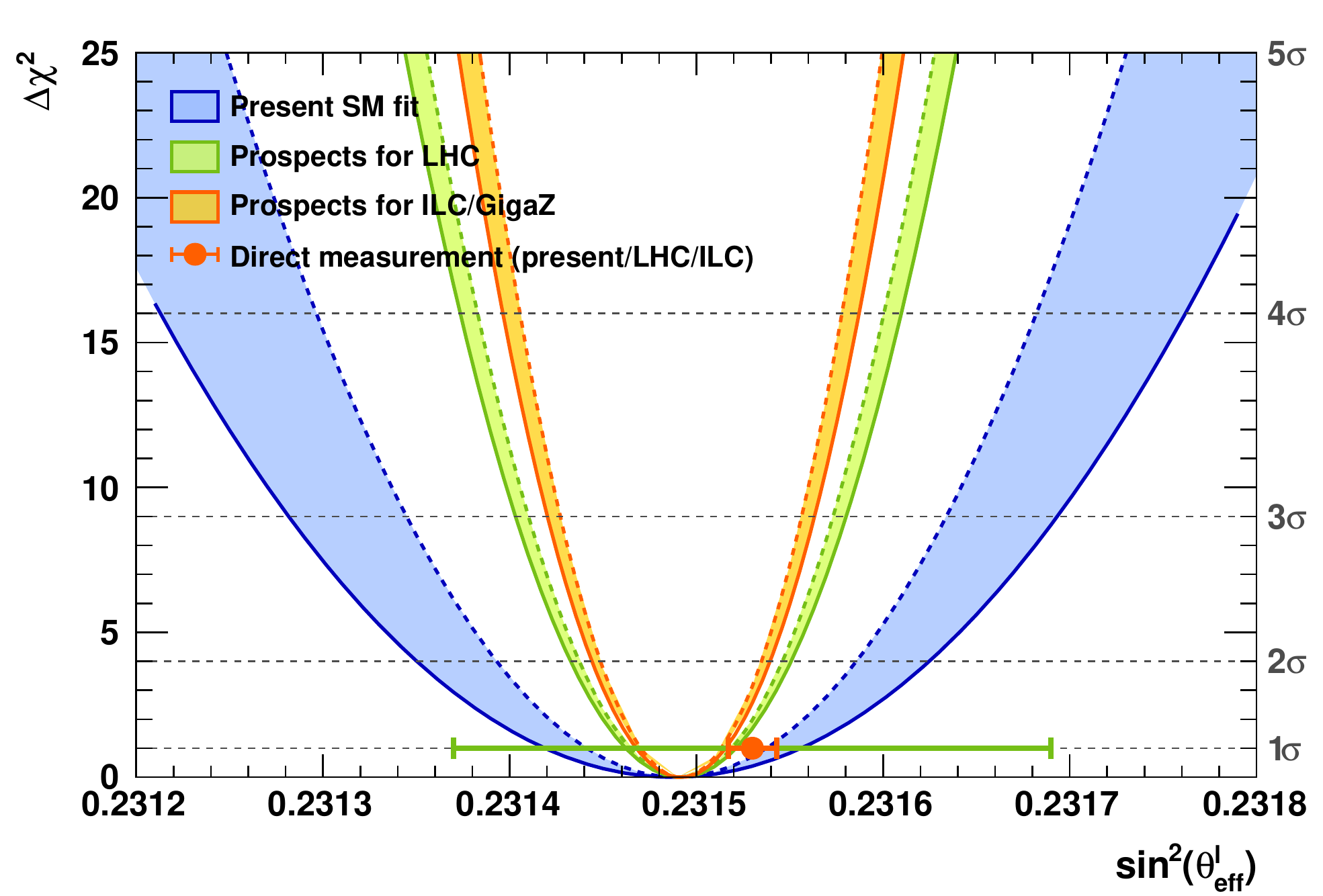}
\vspace{-0.1cm}
\end{center}
\caption[]{Profiles of $\DeltaChi$ versus $M_H$ (top), $M_W$ (middle) and $\sinleff$ (bottom). In blue
 the present result, and in light blue, green and orange the present, LHC and
 ILC/GigaZ scenarios, respectively, all using the future fit setup (reproducing
 $M_H\simeq125\;$GeV) with corresponding uncertainties.
 The impact of the theoretical uncertainties is illustrated by the width of
 the coloured curves. See Table~\ref{tab:results_best}
 for the numerical results of these fits. }
\label{fig:mh_scan}
\end{figure}

Examples of $\Delta\chi^2$ profiles for three key observables are shown in Fig.~\ref{fig:mh_scan}. 
Throughout this section, blue, green and orange curves indicate the present,
future LHC and future ILC/GigaZ scenarios. 
The impact of the theoretical uncertainties is illustrated by the width of 
each coloured curve. 
The light blue curve in the top panel 
in Fig.~\ref{fig:mh_scan} indicates the $M_H$ constraint using the present precision, but 
with the central experimental values adjusted to the future scenarios. 
It allows a direct comparison with the present uncertainties, which depend on the value of $M_H$.

If the extrapolated precision on $M_W$ and $m_t$ can be realised,
the LHC will significantly improve the indirect constraint on $M_H$ (present at
$M_H\simeq125\;$GeV: $^{+33}_{-27}$\;GeV, LHC: $^{+21}_{-18}$\;GeV).  
An even more substantial improvements is expected for the ILC/GigaZ with an
expected uncertainty of $^{+7.4}_{-7.0}$\;GeV.\footnote{If the experimental input
data, currently predicting $M_H=94^{\,+25}_{\,-22}\;\gev$, are left
unchanged with respect to the present central values but had
uncertainties according to the future expectations, a precision of
$^{+16}_{-14}\;\gev$ and $^{+5.6}_{-5.3}$\;GeV is obtained for LHC and ILC/GigaZ respectively.  
A deviation of the measured $M_H$ at a level of $5\sigma$ could  be
established with the ILC/GigaZ fit.}

Correspondingly, the prediction of $M_W$ from the fit (see middle panel of
Fig.~\ref{fig:mh_scan}) can be improved by the LHC
(reduced uncertainty from currently $7.8$\;MeV to $5.5$\;MeV, owing also to
the reduced theoretical uncertainties) and by
the ILC/GigaZ ($2.3$\;MeV). Also shown on the figure are the current
and expected future direct measurements, keeping the central value
unchanged.  A powerful SM test is obtained, confronting measurement
and prediction of $M_W$ at the level of 0.05
per mill.

The prediction of $\sinleff$ from the fit (bottom panel of
Fig.~\ref{fig:mh_scan}) is significantly improved in the LHC and ILC/GigaZ
scenarios, also owing to the improved theoretical precision. The
total uncertainty reduces from currently $6.6 \cdot 10^{-5}$ by
almost a factor of three at the ILC/GigaZ.  Again the current and
expected future direct measurements are also indicated on the figure,
keeping the central value unchanged. No improvement in the precision of
the direct measurement is expected from the LHC, leaving the direct
measurement a factor 5 less precise than the indirect determination.
Only with in the ILC/GigaZ scenario a similar precision between the
prediction and direct measurement can be achieved.

\begin{figure}[p]
\begin{center}
\includegraphics[width=0.73\textwidth]{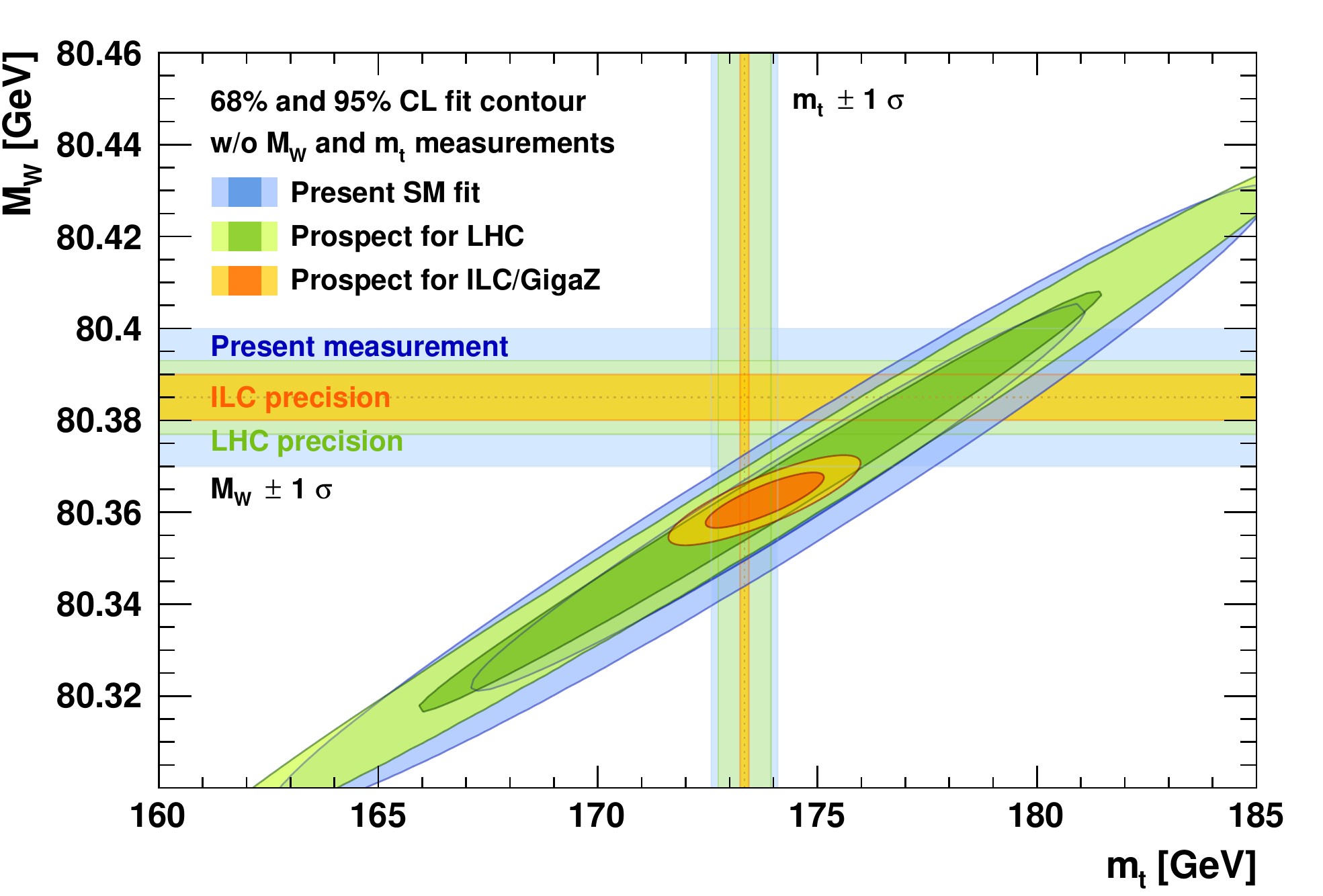}

\vspace{0.4cm}
\includegraphics[width=0.73\textwidth]{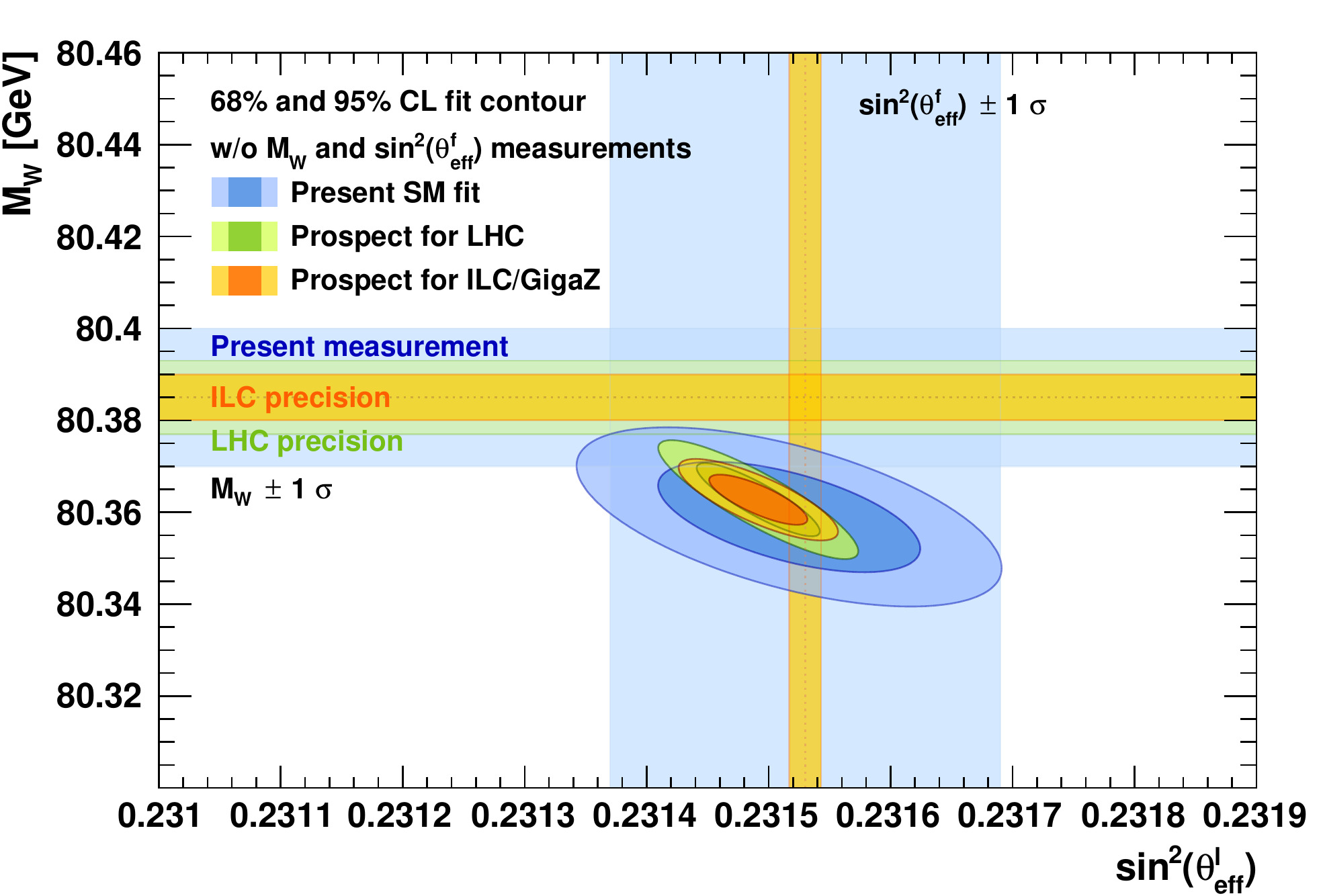}
\end{center}
\vspace{-0.1cm}
\caption[]{ Fit constraints for the present and extrapolated future scenarios
  compared to the direct measurements for the observable pairs $M_W$ versus $m_t$ (top)
  and $M_W$ versus $\sinleff$ (bottom).
  The direct measurements are not included as input measurements in
  the fits. For the future scenarios the central values of the other
  input measurements are adjusted to reproduce the SM with
  $M_H\simeq125\;$GeV. The horizontal and
  vertical bands indicate in blue today's precision of the direct
  measurements, and in light green and orange the extrapolated
  precisions for the LHC and ILC/GigaZ, respectively.
  The ellipses receive significant contributions from the theoretical 
  uncertainties parametrised by $\deltatheo M_W$ and $\deltatheo \sinfeff$.
  For better visibility the measurement ellipses corresponding to two degrees of
  freedom are not drawn.
}
\label{fig:W_vs_top}
\end{figure}

Figure~\ref{fig:W_vs_top} shows the allowed areas obtained for fits with fixed 
variable pairs $M_W$ versus $m_t$ (top) and $M_W$ versus \sinleff (bottom)
in the three scenarios.  
The horizontal and vertical bands display 
the 1$\sigma$ ranges of the current direct measurements (blue), as well as the 
LHC (green) and ILC/GigaZ (orange) expectations in precision. 
A modest improvement in precision is achieved for the LHC, represented by the 
green ellipses, when confronting the direct measurements with the SM predictions.
A much stronger increase in precision and sensitivity is obtained with the ILC/GigaZ
(orange ellipses).

\subsection*{Impact of the individual uncertainties}

\begin{table}[t!]
\setlength{\tabcolsep}{0.0pc}
{\normalsize
\begin{tabular*}{\textwidth}{@{\extracolsep{\fill}}l c c c c@{\hskip 12pt} | c c c c c c} 
\hline\noalign{\smallskip}
\multicolumn{5}{c|}{}                               &   \multicolumn{6}{c}{Experimental uncertainty source [{\footnotesize $\pm 1 \sigma$}]}  \\
Parameter         & $\delta_{\rm meas}$  &   $\delta_{\rm fit}^{\rm tot}$ &  $\delta_{\rm fit}^{\rm theo}$ & $\delta_{\rm fit}^{\rm exp}$ &  $\delta M_W$ &    $\delta M_Z$  &  $\delta m_t$  & $ \delta \sinfeff$ \hspace{-0.2cm}  &    $\delta \Delta\alpha_{\rm had}$ \hspace{-0.2cm}     & $\delta\as$ \hspace{-0.2cm} \\
\noalign{\smallskip}\hline
\multicolumn{11}{c}{\small Present uncertainties}  \\
\hline\noalign{\smallskip}
$ M_{H}$        {\ft [GeV]}            &  $0.4$  & $_{-27}^{+33}$  &  $_{-8}^{+10}$  & $_{-26}^{+31}$  & $_{-23}^{+28}$  & $_{-4}^{+5}$  & $_{-7}^{+10}$  & $_{-23}^{+29}$  & $_{-5}^{+7}$  & $_{-3}^{+4}$ \\
$ M_{W}$        {\ft [MeV]}            &  15     &    $ 7.8$       &  $ 5.0$         & $ 6.0$          & --   & $ 2.5$  & $ 4.3$  & $ 5.1$  & $ 1.6$  & $ 2.5$ \\   
$ M_{Z}$        {\ft [MeV]}            &  2.1    &   $ 12.0$       &  $ 3.7$         & $ 11.4$         & $ 10.5$ & --  & $ 3.5$  & $ 11.2$  & $ 2.2$  & $ 1.4$ \\
$ m_{t}$        {\ft [GeV]}            &  0.8    &   $ 2.5$        &  $ 0.6$         & $ 2.4$          & $ 2.3$  & $ 0.4$ & --  & $ 2.3$  & $ 0.5$  & $ 0.6$ \\ 
$ \sinleff$ $^{(\circ)}$               &  16     &   $ 6.6$        &  $ 4.9$         & $ 4.5$          & $ 3.7$  & $ 1.2$  & $ 2.0$ & --  & $ 3.4$  & $ 1.2$ \\ 
$ \Delta\alpha_{\rm had}$ $^{(\circ)}$ &  10     &    $ 44$        &  $ 13$          & $ 42$           & $ 31$  & $ 6$  & $ 10$  & $ 41$ & --  & $ 2$ \\
\noalign{\smallskip}\hline
\multicolumn{11}{c}{\small LHC prospects}  \\
\hline\noalign{\smallskip}
$ M_{H}$        {\ft [GeV]}            &  $<0.1$ & $_{-18}^{+21}$  & $_{-3}^{+4}$  & $_{-18}^{+20}$  &  $_{-14}^{+17}$  & $_{-5}^{+6}$  & $_{-7}^{+8}$  & $_{-16}^{+18}$  & $_{-2}^{+3}$  & $_{-4}^{+5}$ \\
$ M_{W}$        {\ft [MeV]}            &  8      &     $ 5.5$      & $ 1.8$        & $ 5.2$          &  --  & $ 2.5$  & $ 3.5$  & $ 4.8$  & $ 0.8$  & $ 2.6$ \\  
$ M_{Z}$        {\ft [MeV]}            &  2.1    &   $ 7.2$        & $ 1.4$        & $ 7.0$          &  $ 6.0$ & --  & $ 2.8$  & $ 5.9$  & $ 0.8$  & $ 1.9$ \\
$ m_{t}$        {\ft [GeV]}            &  0.6    & $ 1.5$          & $ 0.2$        & $ 1.5$          &  $ 1.3$  & $ 0.4$ & --  & $ 1.2$  & $ 0.2$  & $ 0.5$ \\
$ \sinleff$ $^{(\circ)}$               &  16     &   $ 3.0$        & $ 1.1$        & $ 2.8$          &  $ 2.5$  & $ 1.1$  & $ 1.4$ & --  & $ 1.5$  & $ 0.9$ \\
$ \Delta\alpha_{\rm had}$ $^{(\circ)}$ & 4.7     & $ 36$           & $ 6$          & $ 36$           &  $ 25$  & $ 9$  & $ 12$  & $ 35$ & --  & $ 5$ \\
\noalign{\smallskip}\hline
\multicolumn{11}{c}{\small ILC/GigaZ prospects}  \\
\hline\noalign{\smallskip}
$ M_{H}$        {\ft [GeV]}            &  $<0.1$  &  $_{-7.0}^{+7.4}$  &  $_{-2.3}^{+2.5}$   & $_{-6.6}^{+6.9}$  & $_{-1.9}^{+3.9}$  & $_{-4.1}^{+4.3}$  & $_{-0.8}^{+0.9}$  & $_{-3.0}^{+3.3}$  & $_{-4.1}^{+4.3}$  & $_{-0.3}^{+0.3}$ \\ 
$ M_{W}$        {\ft [MeV]}            &  5       &    $ 2.3$          &  $ 1.3$             & $ 1.9$            & --  & $ 1.7$  & $ 0.3$  & $ 1.3$  & $ 0.7$  & $ 0.3$ \\  
$ M_{Z}$        {\ft [MeV]}            & 2.1      &   $ 2.7$           &  $ 1.0$             & $ 2.6$            & $ 2.5$ & --  & $ 0.4$  & $ 1.3$  & $ 1.9$  & $ 0.2$ \\
$ m_{t}$        {\ft [GeV]}            &  0.1     & $ 0.8$             &  $ 0.2$             & $ 0.7$            & $ 0.6$  & $ 0.5$ & --  & $ 0.3$  & $ 0.4$  & $ 0.2$ \\
$ \sinleff$ $^{(\circ)}$               &  1.3     &  $ 2.3$            &  $ 1.0$             & $ 2.0$            & $ 1.7$  & $ 1.2$  & $ 0.2$ & --  & $ 1.5$ & 0.1  \\ 
$ \Delta\alpha_{\rm had}$ $^{(\circ)}$ & 4.7      & $ 6.4$             &  $ 3.0$             & $ 5.6$            & $ 2.7$  & $ 4.1$  & $ 0.8$  & $ 3.9$ & --  & $ 0.2$ \\
\noalign{\smallskip}\hline
\noalign{\smallskip}
\end{tabular*}
{\ft
\vspace{-0.3cm}
$^{(\circ)}$In units of $10^{-5}$.
$^{(\star)}$In units of $10^{-4}$
}} \\
\caption{Contributions from the individual experimental and theoretical uncertainty 
         sources to the total uncertainty in the indirect determination of 
         a given observable by the electroweak fit for the three scenarios
         (present, future LHC, ILC/GigaZ). The uncertainty due to $M_H$ is negligible 
         compared to the other observables and is not shown. See text for further 
         discussion. \label{tab:results_errors} }
\end{table} 
Table~\ref{tab:results_errors} shows a breakdown of the predicted
uncertainties of various parameters as obtained from the reduced
electroweak fit for the present and future scenarios.  
The present and prospective experimental precision of the direct measurement, $\delta_{\rm meas}$, 
are given in column two. In column three the total uncertainty from the indirect 
determination, ie. the result from a fit without using the experimental observable of that row, 
$\delta_{\rm fit}^{\rm tot}$, is given. The contributions from the 
theoretical uncertainties, $\delta_{\rm fit}^{\rm theo}$,
and experimental uncertainties, $\delta_{\rm fit}^{\rm exp}$, are shown in columns four and five. 
Columns six to eleven give the uncertainties of the indirect fit determination resulting 
from the experimental uncertainties of the observables listed in the respective columns. 
These uncertainties are obtained as the difference between the result obtained from the full
fit and the result when excluding the experimental uncertainty given in that column.
In this approach the correlations between the fit parameters are
neglected, such that the individual experimental uncertainties do not
add up in quadrature to the full experimental uncertainty as obtained from the fit.
The given individual uncertainties thus show the precision that can be
gained by improving the constraints from a single measurement.

One notices that the dominant uncertainty contributions vary between the three scenarios.
For $M_H$, the precision on the indirect determination is presently
dominated by the uncertainty on the measurements of 
$\sinfeff$ and $M_W$, which does not change for the LHC scenario. For the
ILC/GigaZ however, the uncertainties on $M_Z$ and $\dalphaHadMZ$
become equally important and a total precision of less than $10$\;GeV can be
achieved.
For $M_W$, improvements in the theoretical uncertainty and on $\delta
m_t$ could lead to a precision of $5.5$~MeV for the LHC
scenario and of $2.3$~MeV for the ILC/GigaZ. This would exceed the present
experimental precision by $60\%$ to $75\%$, respectively.
For $\sinleff$, improvements in the theoretical uncertainty and in
$\dalphaHadMZ$ and $m_t$ are expected, and could lead to a precision
on $\sinleff$ of $3.0 \cdot 10^{-5}$ for the future LHC scenario,
which would exceed the present experimental precision by more than a factor of five.
The ILC/GigaZ would rectify the imbalance in precision: a precision of
$1.3 \cdot 10^{-5}$ for the direct measurement would confront an
indirect determination with $2.3 \cdot 10^{-5}$ total uncertainty.

At the ILC/GigaZ a comparable precision between
direct determination and fit constraint would be reached for $M_Z$ and
$\dalphaHadMZ$, owing to the improved precision on $M_W$ and
$\sinfeff$. Also, an indirect constraint on $m_t$ of 1\;GeV would
be possible.

Independently from the other improvements, the determination of $\asZ$ from $R_\l^0$ would 
also greatly benefit from the ILC/GigaZ. The current $\asZ$ precision of $30\cdot10^{-4}$
is dominated by experimental uncertainties and could be improved to $9\cdot10^{-4}$. 
It would be the most precise experimental determination of the strong coupling constant, only
challenged by calculations from Lattice QCD.

\subsection*{Prospects for the oblique parameter determination}

\begin{figure}[tp]
\begin{center}
\includegraphics[width=0.73\textwidth]{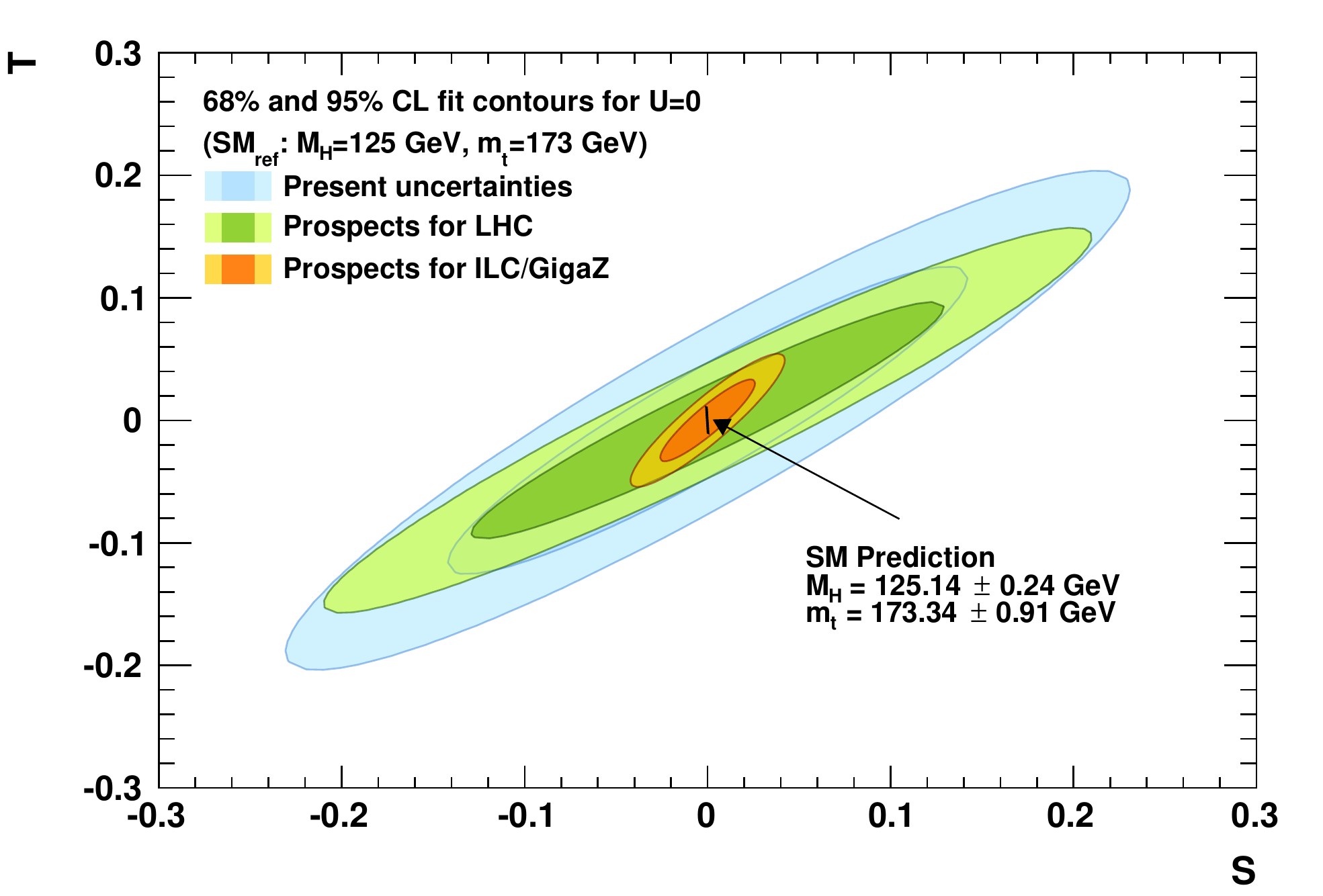}
\end{center}
\vspace{-0.1cm}
\caption[]{Comparison of the present (light blue), the LHC
  (green) and the ILC/GigaZ prospects (orange)
  on the oblique parameters
  $S$ and $T$, with the $U$ parameter fixed to zero.
  The shift in the position of
  the ellipses between the present data and future scenarios is caused 
  by the different central values used for the electroweak observables 
  in these scenarios. The future scenarios are by construction centred 
  at $S=T=0$. The SM prediction within uncertainties is indicated by the 
  thin black stroke.}
\label{fig:STU_future}
\end{figure}

The expected future constraints on $S$ and $T$ for a fixed 
value of $U=0$ are shown in Fig.~\ref{fig:STU_future}. 
The results from the fit of the present 
scenario with central values adjusted to obtain $M_H\simeq125\;$GeV
are shown in light blue.
The shift in the central values between the light blue ellipse and
the results shown in Fig.~\ref{fig:STU} originate 
from the different
central values used for the electroweak observables. 
By construction the ellipses are centred around $S=T=0$. 
The uncertainties in the present scenario are larger by about 0.01 in
$S$ and 0.02 in $T$ due to the reduced list of observables used in the prospective fit, 
as discussed in the beginning of this section. 

Compared to to the present scenario 
only a minor improvement is expected for the
LHC scenario. A reduction of the uncertainty by a factor of three to four
is however expected for the ILC/GigaZ.
The numerical values of the uncertainties on $S$ and $T$ are given in 
Table~\ref{tab:results_best}. 
The parameters $S$ and $T$ are strongly correlated, with correlation
coefficients of $0.93$, $0.96$ and $0.91$ for the present, LHC and
ILC/GigaZ scenarios.

Additional variables like the total width of the $Z$,
$\Gamma_Z$, which could be measured to an accuracy of $0.8$~MeV at
the ILC/GigaZ~\cite{ILCTDR13}, improve the precision on $\delta S$
and $\delta T$ by about 10\%.

\section{Status and prospects for the Higgs couplings determination}
\label{sec:higgscouplings}

To test the validity of the SM and look for signs of new physics, 
precision measurements of the properties of the Higgs boson are of critical importance. 
Key are the couplings to the SM fermions and bosons, which are predicted 
to depend linearly on the fermion mass and quadratically on the boson mass.

\begin{figure}[tp]
\begin{center}
\includegraphics[width=0.7\textwidth]{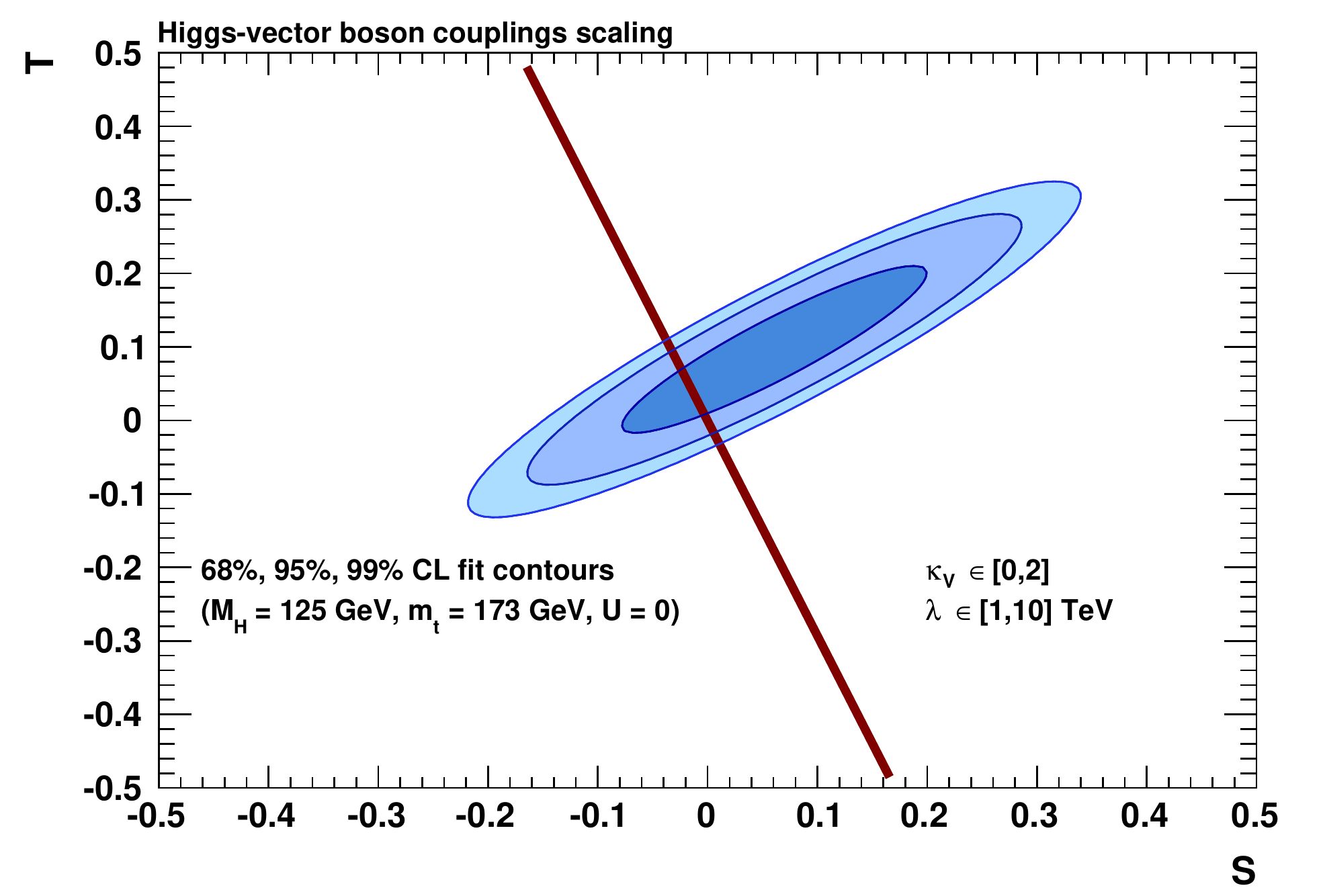} 

\vspace{0.3cm}
\includegraphics[width=0.7\textwidth]{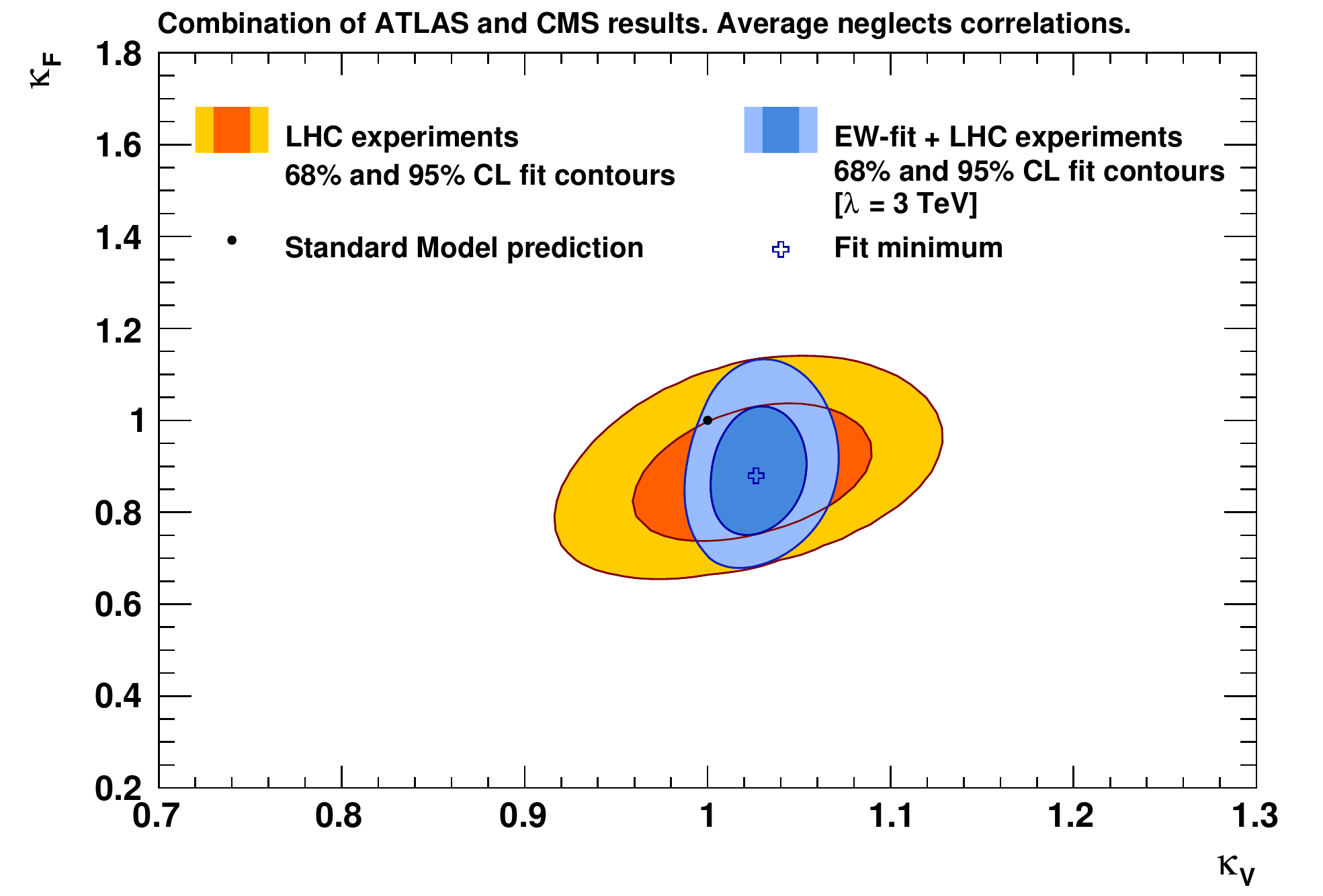}
\end{center}
\vspace{-0.1cm}
  \caption[]{
Top: Contour lines of 68\% and 95\% CL allowed regions for fixed values of $S$ and $T$ with $U=0$ for the 
present data (blue). Overlaid (dark red) is the predicted line for $S$ and $T$ for $\kappa_V \in [0,2]$ and $\lambda \in [1,10]$~TeV.
Bottom: Measurement of $\kappa_F$ versus $\kappa_V$ at 68\% and 95\% CL from a private combination of present ATLAS and CMS results 
(orange), overlaid with the constraint of $\kappa_F$ versus $\kappa_V$ when including the EW-fit (blue).}
\label{fig:current_mw_vs_cv}
\end{figure}

Modified Higgs couplings have been probed by ATLAS and CMS in various benchmark models~\cite{atlashc,atlastautau,TheATLAScollaboration:2013lia,cmshc,Chatrchyan:2013iaa,Chatrchyan:2013mxa,Chatrchyan:2014nva,Chatrchyan:2013zna}. 
These employ an effective theory approach, where higher-order modifiers to a phenomenological 
Lagrangian are matched at tree-level to the SM Higgs boson couplings.   
In one popular model all boson and all fermion couplings are modified in the same way,
scaled by the constants $\kappa_V$ and $\kappa_F$, respectively, where $\kappa_V=\kappa_F=1$ for 
the SM.\footnote{Equivalent notations are: $\kappa_V\equiv c_V \equiv a$, and $\kappa_F\equiv c_F \equiv c$.}
This benchmark model uses the explicit assumption that no other new physics is present,
e.g., there are no additional loops in the production or decay of the Higgs boson,
and no invisible Higgs decays and undetectable contributions to its decay width.
For details see Ref.~\cite{ref:hxswg}.

The combined analysis of electroweak precision data and Higgs signal-strength 
measurements has been studied by several 
groups~\cite{ref:espinoza,ref:1205.0011,Eberhardt:2012np,ref:1209.6382,ref:1211.4580,Falkowski:2013dza,ref:1306.4644}.
The main effect of this model on the electroweak precision observables  is from the 
modified Higgs coupling to gauge bosons, and manifests itself through loop diagrams involving 
the longitudinal degrees of freedom of these bosons.
The corrections to the $Z$ and $W$ boson propagators 
can be expressed in terms of the \ST parameters~\cite{ref:espinoza},
\begin{equation}
S=\frac{1}{12\pi}(1-\kappa_V^2) \ln\!\frac{\Lambda^{\!2}}{M_H^2}\:,\quad T=-\frac{3}{16\pi \cosleff }(1-\kappa_V^2) \ln\! \frac{\Lambda^{\!2}}{M_H^2}\:,\quad \Lambda = \frac{\lambda}{\sqrt{|1-\kappa_V^2|}}\:,
\end{equation}
and $U=0$. 
The cut-off scale $\Lambda$ represents the mass scale of the new states that unitarise 
longitudinal gauge-boson scattering, as required in this model.
Note that the less $\kappa_V$ deviates from one, the higher the scale of new physics.
Most BSM models with additional Higgs bosons giving positive corrections to the $W$ mass predict values of $\kappa_V$ smaller than $1$.
Here the nominator $\lambda$ is varied between $1$ and $10$ TeV, and is nominally fixed 
to $3$ TeV ($4\pi v$).

Figure~\ref{fig:current_mw_vs_cv}~(top) shows the predictions for $S$ and $T$, 
profiled over $\kappa_V$ and $\lambda$, together with the allowed 
regions for $S$ and $T$ from the current electroweak fit.
The length of the predicted line covers a variation in $\kappa_V$ between $[0,2]$, the width covers the variation in $\lambda$.

The bottom panel of Fig.~\ref{fig:current_mw_vs_cv} shows $\kappa_V$ and $\kappa_F$ as obtained from 
a private combination of ATLAS and CMS results using all publicly available information on the measured Higgs
signal strength modifiers $\mu_i$. 
Also shown is the combined constraint on $\kappa_V$ (and $\kappa_F$) from the LHC 
experiments and the electroweak fit.

The published Higgs coupling measurements of $\mu_{\rm ggF+ttH}$ versus $\mu_{\rm VBF+VH}$ from ATLAS and CMS used in this combination are summarised in Table~\ref{tab:cmsmuis}.
The measurements from the ATLAS Higgs to di-boson channels are published likelihood scans~\cite{atlashc}.
The CMS results in Table~\ref{tab:cmsmuis} are approximate values derived from public likelihood iso-contour lines.
Correlations of the theory and detector related uncertainties between the various $\mu_i$ are neglected in the combination, 
as these are not provided by the experiments.
We find that the individual experimental combinations of ATLAS and CMS for $\kappa_V$ (and $\kappa_F$) are approximately reproduced by this simplified procedure.
The measured values from this combination are $\kappa_V = 1.026^{\,+0.042}_{\,-0.044}$ 
and $\kappa_F = 0.88^{\,+0.10}_{\,-0.09}$.

\begin{table}[t]
\setlength{\tabcolsep}{0.0pc}
{\normalsize                                                                                   
\begin{tabular*}{\textwidth}{@{\extracolsep{\fill}}llcccl} 
\hline\noalign{\smallskip}
Experiment & Channel  &  $\mu_{\rm ggF+ttH}$ & $\mu_{\rm VBF+VH}$ & Correlation & Ref. \\
\noalign{\smallskip}\hline\noalign{\smallskip}
ATLAS       & $H\to\gamma\gamma,\;WW^{\star}, \;ZZ^\star$      &  \multicolumn{3}{c}{Published 2D-likelihood scan}  & \cite{atlashc}  \\
\noalign{\smallskip}\hline\noalign{\smallskip}
       & $H\to\gamma\gamma$      & $1.13^{+0.37}_{-0.31}$      &  $1.15^{+0.63}_{-0.58}$  &  $-0.45$ &   \cite{Khachatryan:2014ira}  \\
       & $H\to WW^{\star}$        & $0.70^{+0.25}_{-0.20}$      &  $0.70^{+0.65}_{-0.50}$  &  $-0.26$ &  \cite{Chatrchyan:2013iaa}   \\
CMS    & $H\to ZZ^\star$               & $0.80^{+0.46}_{-0.36}$      &  $1.70^{+2.20}_{-2.10}$  &  $-0.75$ &   \cite{Chatrchyan:2013mxa}  \\
       & $H\to \tau\tau$  & $0.50^{+0.53}_{-0.53}$      &  $1.30^{+0.46}_{-0.40}$  &  $-0.40$ &  \cite{Chatrchyan:2014nva}   \\
       & $H\to bb$    & -- & $1.00^{+0.50}_{-0.50}$ & -- & \cite{Chatrchyan:2013zna} \\
\noalign{\smallskip}\hline
\noalign{\smallskip}
\end{tabular*} \\
}
\caption{The ATLAS and CMS Higgs coupling measurements of $\mu_{\rm ggF+ttH}$ and 
         $\mu_{\rm VBF+VH}$, and their correlations, as used in this study. Unless 
         where available, the central values, uncertainties and correlations have 
         been estimated from published or public likelihood iso-contour lines.}
\label{tab:cmsmuis} 
\end{table}
\begin{figure}[tp]
\begin{center}
\includegraphics[width=0.73\textwidth]{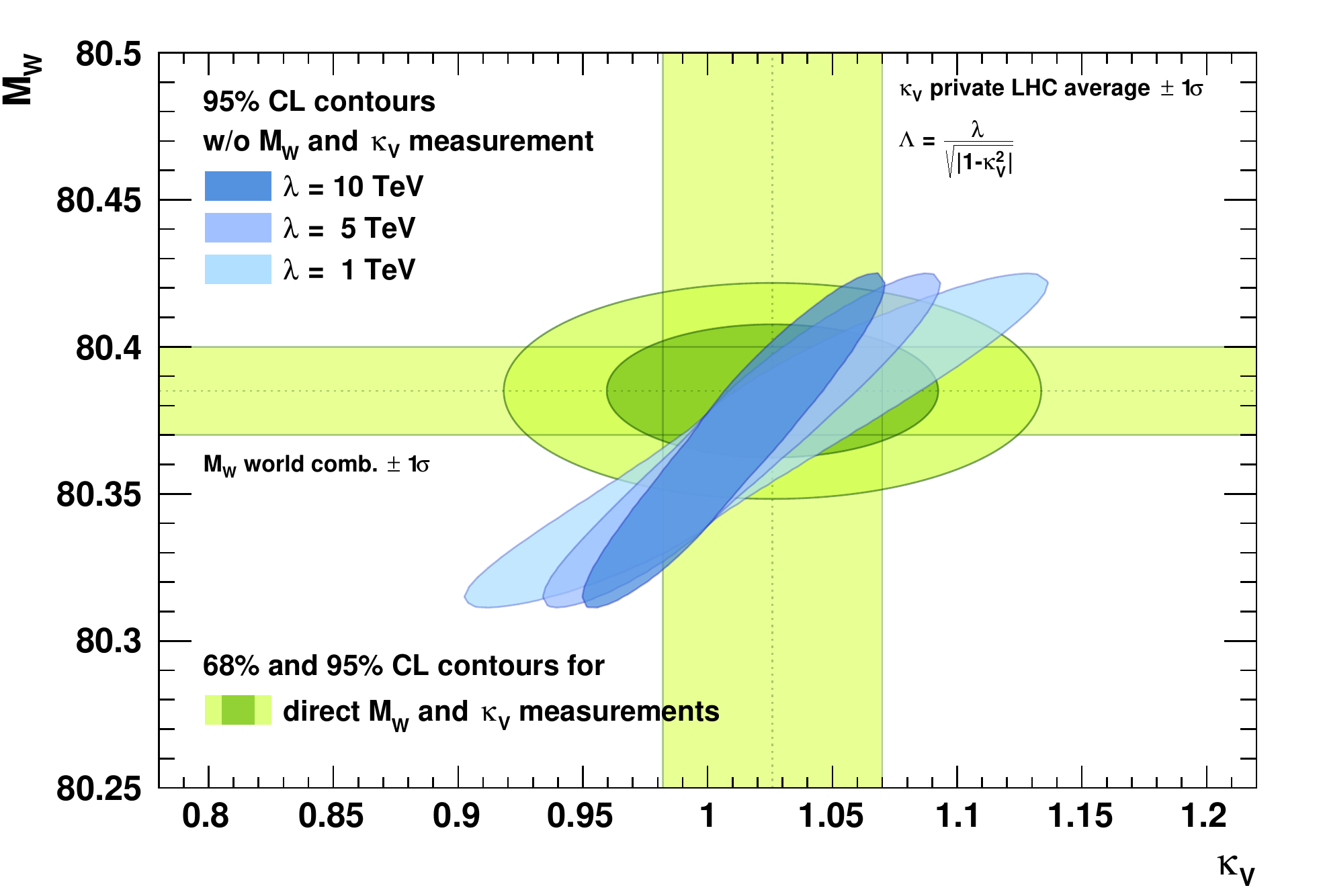} 

\vspace{0.3cm}
\includegraphics[width=0.73\textwidth]{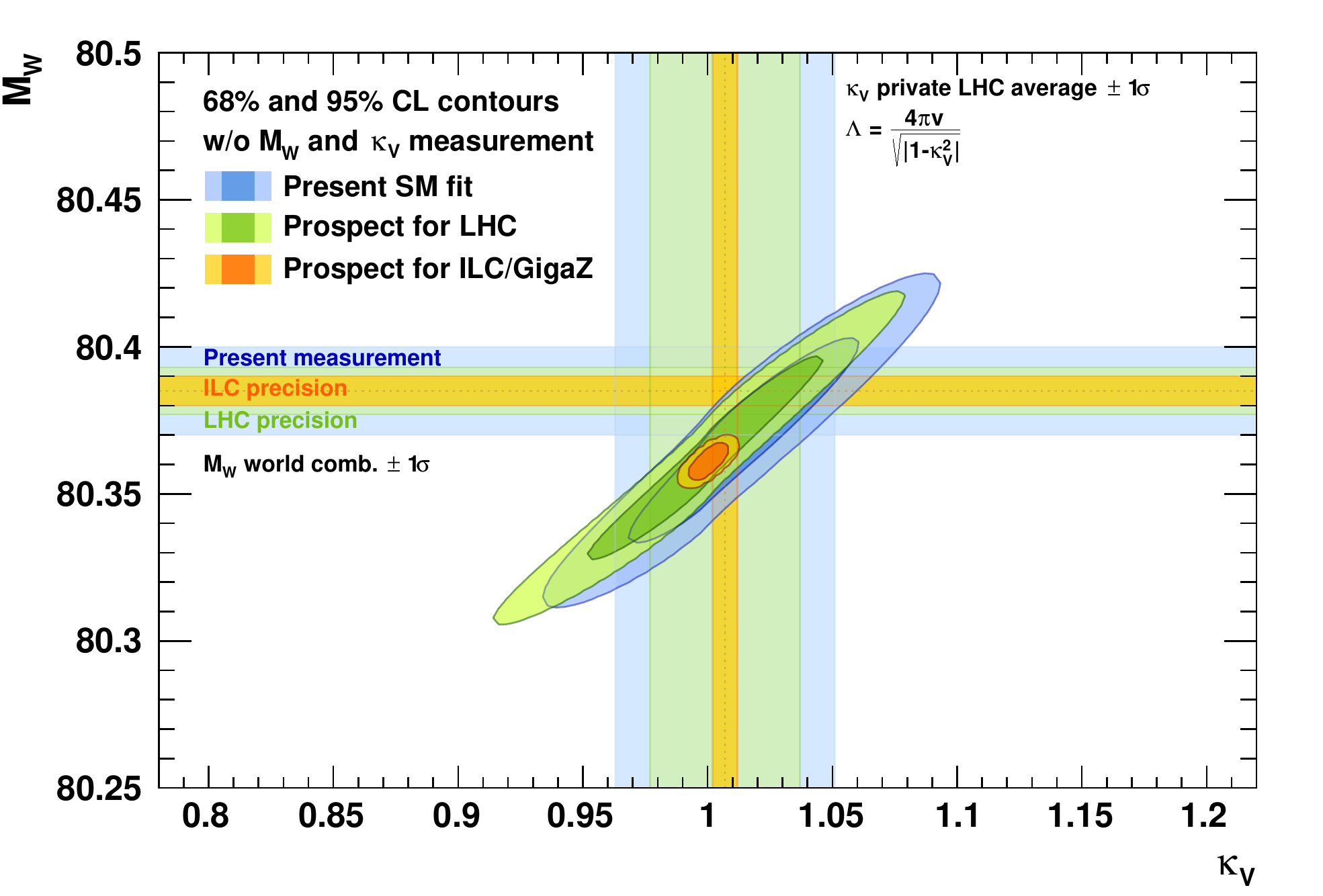} 
\end{center}
\vspace{-0.1cm}
 \caption[]{
Top: Comparison of the direct $M_W$ and $\kappa_V$ measurements
(horizontal and vertical green bands) with the contours of 95\% CL allowed regions obtained from global fits for various values of the 
cut-off scale $\lambda$, in which the direct measurements of $M_W$ and $\kappa_V$ are 
not included.
Bottom: Similar comparison of the direct $M_W$ and $\kappa_V$ measurements and their indirect predictions 
for $\lambda=3$~TeV, for the present (blue) and the ILC/GigaZ (yellow/orange) precision, 
at 68\% and 95\% CL. For better visibility the experimental ellipse is not drawn in the lower plot. 
}
\label{fig:hcouplfuture}
\end{figure}

The electroweak fit results in $\kappa_V=1.037^{+0.029}_{-0.026}$, $1.027^{+0.020}_{-0.019}$, and $1.021^{+0.015}_{-0.014}$, for 
cut-off parameters $\lambda = 1$\;TeV, 3\;TeV and 10\;TeV, respectively, where $\lambda$ has been 
fixed during each of the fits. 
Including constraints from electroweak precision observables, the constraint on $\kappa_V$ 
can be improved by a factor of more than three.
There is a mild dependence -- both in the central value and uncertainty -- on the chosen value 
for $\lambda$, but all values result in small but positive deviations from unity.
For $\kappa_V\sim 1.03$ and $\lambda = 4\pi v$, the new physics scale is $\Lambda \gtrsim 13$~TeV.

The slight positive deviation of $\kappa_V$ from $1$ is driven by the small discrepancy between the 
observed and predicted values of the $W$ mass, as shown in Fig.~\ref{fig:hcouplfuture}~(top).
The figure exhibits the strong correlation between the two quantities,
and also the dependence on the chosen value of $\lambda$.
To determine the predicted ellipses, the measured value of $M_W$ and the 
current measurements of $\mu_i$ have been removed from the EW fit.

Figure~\ref{fig:hcouplfuture}~(bottom) shows the prospects for predicting and measuring $\kappa_V$ versus 
$M_W$ at the LHC and ILC/GigaZ. 
For LHC, the predicted precision on $\kappa_V$ is largely limited by theoretical uncertainties
somewhat optimistically set to $3\%$~\cite{atlashcpred,cmshcpred}.
For the ILC, the predicted uncertainties on the measurements of the Higgs to $W$ and $Z$ gauge boson coupling constants are both $1\%$~\cite{ref:peskin}.
Assuming custodial symmetry, these uncertainties have been averaged in the figure.
For the indirect LHC and ILC predictions, the central values of the 
electroweak observables have been shifted 
to match the Higgs mass of $125$~GeV, with $\kappa_{V} = 1$.
The nominal value of $\lambda$ is $3$~TeV. 
Varying $\lambda$ between $1$\;TeV, 3\;TeV and 10\;TeV, the central value of $\kappa_{V}$ remains 
unchanged at $1$, but its uncertainty varies between $0.008$ and $0.015$ at the LHC 
and between $0.003$ and $0.005$ for the ILC scenario. 
The numbers obtained for $\lambda = 3$\;TeV are summarised in Table~\ref{tab:results_best}.
Assuming  the present central values of $\kappa_V$ and $M_W$, 
the deviation 
of $\kappa_V$ from one would become significant.

\section{Conclusion}

We have updated in this paper the results from the global electroweak fit
using full fermionic two-loop calculations for the partial widths
and branching ratios of the $Z$ boson~\cite{Freitas:2014hra}, and including
a detailed assessment of the impact of theoretical uncertainties.
The prospects of the fit in view of future colliders, 
namely the Phase-1 LHC and the ILC with GigaZ mode, were also studied. 
Significant increase in the predictive power of the fit
was found in both scenarios, where in particular the ILC/GigaZ
provides excellent sensitivity to indirect new physics.
We have also carried out an analysis of the Higgs coupling data in a
benchmark model with modified effective SM Higgs couplings to
fermions and bosons parametrised by one parameter each. The inclusion of
electroweak precision observables yields constraints on the bosonic
coupling $\kappa_V$ that are about twice stronger than current Higgs coupling
data alone, while the precision on the fermionic coupling $\kappa_F$
is not improved.

\subsubsection*{Acknowledgements}
\label{sec:Acknowledgments}

\begin{details}
  We thank Ayres Freitas for intensive discussions, help concerning
  the electroweak calculations, and a careful reading of this 
  manuscript. This work is supported by
  the German Research Foundation (DFG) in the Collaborative Research
  Centre (SFB) 676 ``Particles, Strings and the Early Universe''
  located in Hamburg.

\end{details}

\addcontentsline{toc}{section}{References}
\bibliography{References}{}

\begin{thebibliography}{10}

\bibitem{LEPEWWG}
ALEPH, DELPHI, L3, OPAL, LEP Electroweak, S.~Schael {\em et~al.},
\newblock Phys.Rept. {\bf 532}, 119 (2013), [1302.3415].

\bibitem{Erler:2010wa}
J.~Erler,
\newblock Phys. Rev. {\bf D81}, 051301 (2010), [1002.1320].

\bibitem{Flacher:2008zq}
H.~Flacher {\em et~al.},
\newblock Eur. Phys. J. {\bf C60}, 543 (2009), [0811.0009],
\newblock Erratum-ibid. C71 (2011) 1718.

\bibitem{Baak:2011ze}
M.~Baak {\em et~al.},
\newblock Eur. Phys. J. {\bf C72}, 2003 (2012), [1107.0975].

\bibitem{ref:1306.4644}
M.~Ciuchini, E.~Franco, S.~Mishima and L.~Silvestrini,
\newblock JHEP {\bf 1308}, 106 (2013), [1306.4644].

\bibitem{ATLASHiggs}
ATLAS Collaboration,
\newblock Phys. Lett. B  (2012), [1207.7214].

\bibitem{CMSHiggs}
CMS Collaboration,
\newblock Phys. Lett. B  (2012), [1207.7235].

\bibitem{Baak:2012kk}
M.~Baak {\em et~al.},
\newblock Eur. Phys. J. {\bf C72}, 2205 (2012), [1209.2716].

\bibitem{Eberhardt:2012np}
O.~Eberhardt {\em et~al.},
\newblock Phys.Rev.Lett. {\bf 109}, 241802 (2012), [1209.1101].

\bibitem{Freitas:2014hra}
A.~Freitas,
\newblock JHEP {\bf 1404}, 070 (2014), [1401.2447].

\bibitem{ILCTDR13}
H.~Baer {\em et~al.},
\newblock 1306.6352.

\bibitem{ALEPH:2005ema}
{The ALEPH, DELPHI, L3, OPAL, SLD Collaborations, the LEP Electroweak Working
  Group, the SLD Electroweak and Heavy Flavour Working Groups},
\newblock Phys. Rept. {\bf 427}, 257 (2006), [hep-ex/0509008].

\bibitem{Awramik:2004ge}
M.~Awramik, M.~Czakon, A.~Freitas and G.~Weiglein,
\newblock Phys. Rev. Lett. {\bf 93}, 201805 (2004), [hep-ph/0407317].

\bibitem{Awramik:2006uz}
M.~Awramik, M.~Czakon and A.~Freitas,
\newblock JHEP {\bf 0611}, 048 (2006), [hep-ph/0608099].

\bibitem{Avdeev:1994db}
L.~Avdeev, J.~Fleischer, S.~Mikhailov and O.~Tarasov,
\newblock Phys. Lett. {\bf B336}, 560 (1994), [hep-ph/9406363],
\newblock {[Erratum-ibid. {\bf{B349}}, 597 (1994)]}.

\bibitem{Chetyrkin:1995ix}
K.~Chetyrkin, J.~H. K{\"{u}}hn and M.~Steinhauser,
\newblock Phys. Lett. {\bf B351}, 331 (1995), [hep-ph/9502291].

\bibitem{Chetyrkin:1995js}
K.~Chetyrkin, J.~H. K{\"{u}}hn and M.~Steinhauser,
\newblock Phys. Rev. Lett. {\bf 75}, 3394 (1995), [hep-ph/9504413].

\bibitem{vanderBij:2000cg}
J.~van~der Bij, K.~Chetyrkin, M.~Faisst, G.~Jikia and T.~Seidensticker,
\newblock Phys. Lett. {\bf B498}, 156 (2001), [hep-ph/0011373].

\bibitem{Faisst:2003px}
M.~Faisst, J.~H. K{\"{u}}hn, T.~Seidensticker and O.~Veretin,
\newblock Nucl. Phys. {\bf B665}, 649 (2003), [hep-ph/0302275].

\bibitem{Schroder:2005db}
Y.~Schr{\"{o}}der and M.~Steinhauser,
\newblock Phys. Lett. {\bf B622}, 124 (2005), [hep-ph/0504055].

\bibitem{Chetyrkin:2006bj}
K.~G. Chetyrkin, M.~Faisst, J.~H. K{\"{u}}hn, P.~Maierhofer and C.~Sturm,
\newblock Phys. Rev. Lett. {\bf 97}, 102003 (2006), [hep-ph/0605201].

\bibitem{Boughezal:2006xk}
R.~Boughezal and M.~Czakon,
\newblock Nucl. Phys. {\bf B755}, 221 (2006), [hep-ph/0606232].

\bibitem{Awramik:2008gi}
M.~Awramik, M.~Czakon, A.~Freitas and B.~Kniehl,
\newblock Nucl. Phys. {\bf B813}, 174 (2009), [0811.1364].

\bibitem{Awramik:2003rn}
M.~Awramik, M.~Czakon, A.~Freitas and G.~Weiglein,
\newblock Phys. Rev. {\bf D69}, 053006 (2004), [hep-ph/0311148].

\bibitem{Freitas:2013dpa}
A.~Freitas,
\newblock Phys. Lett. {\bf B730}, 50 (2014), [1310.2256].

\bibitem{Freitas:2012sy}
A.~Freitas and Y.-C. Huang,
\newblock JHEP {\bf 1208}, 050 (2012), [1205.0299],
\newblock {Erratum-ibid. {\bf 1305} (2013) 074, Erratum-ibid. {\bf 1310} (2013)
  044}.

\bibitem{Chetyrkin:1996ia}
K.~Chetyrkin, J.~H. Kuhn and A.~Kwiatkowski,
\newblock Phys. Rept. {\bf 277}, 189 (1996).

\bibitem{Baikov:2008jh}
P.~Baikov, K.~Chetyrkin and J.~H. Kuhn,
\newblock Phys. Rev. Lett. {\bf 101}, 012002 (2008), [0801.1821].

\bibitem{Baikov:2012er}
P.~Baikov, K.~Chetyrkin, J.~Kuhn and J.~Rittinger,
\newblock Phys. Rev. Lett. {\bf 108}, 222003 (2012), [1201.5804].

\bibitem{Kataev:1992dg}
A.~Kataev,
\newblock Phys. Lett. {\bf B287}, 209 (1992).

\bibitem{Czarnecki:1996ei}
A.~Czarnecki and J.~H. K{\"{u}}hn,
\newblock Phys. Rev. Lett. {\bf 77}, 3955 (1996), [hep-ph/9608366].

\bibitem{Harlander:1997zb}
R.~Harlander, T.~Seidensticker and M.~Steinhauser,
\newblock Phys. Lett. {\bf B426}, 125 (1998), [hep-ph/9712228].

\bibitem{Cho:2011rk}
G.-C. Cho, K.~Hagiwara, Y.~Matsumoto and D.~Nomura,
\newblock JHEP {\bf 1111}, 068 (2011), [1104.1769].

\bibitem{Hoang:2008yj}
A.~H. Hoang, A.~Jain, I.~Scimemi and I.~W. Stewart,
\newblock Phys. Rev. Lett. {\bf 101}, 151602 (2008), [0803.4214].

\bibitem{Hoang:2008xm}
A.~H. Hoang and I.~W. Stewart,
\newblock Nucl. Phys. Proc. Suppl. {\bf 185}, 220 (2008), [0808.0222].

\bibitem{Buckley:2011ms}
A.~Buckley {\em et~al.},
\newblock Phys.Rept. {\bf 504}, 145 (2011), [1101.2599].

\bibitem{Moch:2014tta}
S.~Moch {\em et~al.},
\newblock 1405.4781.

\bibitem{Skands:2007zg}
P.~Skands and D.~Wicke,
\newblock Eur. Phys. J. {\bf C52}, 133 (2007), [hep-ph/0703081].

\bibitem{Wicke:2008iz}
D.~Wicke and P.~Z. Skands,
\newblock Nuovo Cim. {\bf B123}, S1 (2008), [hep-ph/0807.3248].

\bibitem{Chetyrkin:1999qi}
K.~Chetyrkin and M.~Steinhauser,
\newblock Nucl. Phys. {\bf B573}, 617 (2000), [hep-ph/9911434].

\bibitem{Melnikov:2000qh}
K.~Melnikov and T.~v. Ritbergen,
\newblock Phys. Lett. {\bf B482}, 99 (2000), [hep-ph/9912391].

\bibitem{top13mangano}
M.~Mangano,
\newblock {\em private communication}, Jun 2014. See also presentation at
  TOP2012 {\em Interpreting the top quark mass: theoretical and MC aspects},
  Nov 2012.

\bibitem{CMS:2012ixa}
CMS Collaboration,
\newblock CMS-PAS-TOP-12-029  (2012).

\bibitem{Hoang:2000yr}
A.~Hoang {\em et~al.},
\newblock Eur. Phys. J. direct {\bf C2}, 1 (2000), [hep-ph/0001286].

\bibitem{Beringer:1900zz}
Particle Data Group, J.~Beringer {\em et~al.},
\newblock Phys. Rev. {\bf D86}, 010001 (2012).

\bibitem{Davier:2010nc}
M.~Davier, A.~Hoecker, B.~Malaescu and Z.~Zhang,
\newblock Eur. Phys. J. {\bf C71}, 1515 (2011), [1010.4180].

\bibitem{ATLAS:2014wva}
ATLAS Collaboration, CDF Collaboration, CMS Collaboration, D0 Collaboration,
\newblock 1403.4427.

\bibitem{Aad:2014aba}
ATLAS Collaboration,
\newblock 1406.3827.

\bibitem{CMS-PAS-HIG-14-009}
CMS Collaboration,
\newblock CMS-PAS-HIG-14-009  (2014).

\bibitem{Peskin:1990zt}
M.~E. Peskin and T.~Takeuchi,
\newblock Phys. Rev. Lett. {\bf 65}, 964 (1990).

\bibitem{Peskin:1991sw}
M.~E. Peskin and T.~Takeuchi,
\newblock Phys. Rev. {\bf D46}, 381 (1992).

\bibitem{Bozzi:2011ww}
G.~Bozzi, J.~Rojo and A.~Vicini,
\newblock Phys. Rev. {\bf D83}, 113008 (2011), [1104.2056].

\bibitem{Hawkings:1999ac}
R.~Hawkings and K.~M{\"o}nig,
\newblock Eur. Phys. J. direct {\bf C1}, 8 (1999), [hep-ex/9910022].

\bibitem{Davier:2012}
M.~Davier,
\newblock {\em private communication}, Nov 2012.

\bibitem{atlashc}
ATLAS Collaboration,
\newblock Phys.Lett. {\bf B} (2013), [1307.1427].

\bibitem{atlastautau}
ATLAS Collaboration,
\newblock ATLAS-CONF-2013-108  (2013).

\bibitem{TheATLAScollaboration:2013lia}
ATLAS Collaboration,
\newblock ATLAS-CONF-2013-079  (2013).

\bibitem{cmshc}
CMS Collaboration,
\newblock CMS-PAS-HIG-13-005  (2013).

\bibitem{Chatrchyan:2013iaa}
CMS Collaboration,
\newblock JHEP {\bf 1401}, 096 (2014), [1312.1129].

\bibitem{Chatrchyan:2013mxa}
CMS Collaboration,
\newblock Phys. Rev. {\bf D89}, 092007 (2014), [1312.5353].

\bibitem{Chatrchyan:2014nva}
CMS Collaboration,
\newblock JHEP {\bf 1405}, 104 (2014), [1401.5041].

\bibitem{Chatrchyan:2013zna}
CMS Collaboration,
\newblock Phys. Rev. {\bf D89}, 012003 (2014), [1310.3687].

\bibitem{ref:hxswg}
LHC Higgs Cross Section Working Group, A.~David {\em et~al.},
\newblock 1209.0040.

\bibitem{ref:espinoza}
J.~Espinosa, C.~Grojean, M.~Muhlleitner and M.~Trott,
\newblock JHEP {\bf 1212}, 045 (2012), [1207.1717].

\bibitem{ref:1205.0011}
M.~Farina, C.~Grojean and E.~Salvioni,
\newblock JHEP {\bf 1207}, 012 (2012), [1205.0011].

\bibitem{ref:1209.6382}
B.~Batell, S.~Gori and L.-T. Wang,
\newblock JHEP {\bf 1301}, 139 (2013), [1209.6382].

\bibitem{ref:1211.4580}
T.~Corbett, O.~Eboli, J.~Gonzalez-Fraile and M.~Gonzalez-Garcia,
\newblock Phys.Rev. {\bf D87}, 015022 (2013), [1211.4580].

\bibitem{Falkowski:2013dza}
A.~Falkowski, F.~Riva and A.~Urbano,
\newblock JHEP {\bf 1311}, 111 (2013), [1303.1812].

\bibitem{Khachatryan:2014ira}
CMS Collaboration,
\newblock 1407.0558.

\bibitem{atlashcpred}
ATLAS Collaboration,
\newblock ATL-PHYS-PUB-2013-007  (2013).

\bibitem{cmshcpred}
CMS Collaboration,
\newblock CMS-NOTE-2012-006  (2012).

\bibitem{ref:peskin}
M.~E. Peskin,
\newblock 1207.2516.

\end{thebibliography}

\end{document}